\documentclass[aps,floatfix,amsmath,nofootinbib,amssymb,superscriptaddress]{revtex4-2}

\usepackage{overpic}
\usepackage{amssymb}
\usepackage{indentfirst}
\usepackage{feynmf}   
\usepackage{slashed}  
\usepackage{cases}
\usepackage{color}
\usepackage{multirow}
\usepackage{epstopdf}
\usepackage{graphicx,color,bm}
\usepackage{epstopdf}
\usepackage{amsmath} 

\usepackage[colorlinks,
            citecolor=green,
            anchorcolor=red,
            menucolor=red,
            linkcolor=red,
            filecolor=red,
            runcolor=red,
            urlcolor=blue,
            frenchlinks=red]{hyperref}

\begin{document}

\title{Near-threshold photon-proton production of $J/\psi$ and $\Upsilon$}

\author{Chentao Tan}
\author{Zhun Lu}
\email[]{zhunlu@seu.edu.cn}
\affiliation{School of Physics, Southeast University, Nanjing 211189, China}

\begin{abstract}
We study the near-threshold exclusive photoproduction of heavy vector mesons (quarkonia $J/\psi$ and $\Upsilon$) off the proton within the framework of generalized parton distribution (GPD) factorization. The gluon GPDs are computed using a spectator model in which the proton emits a gluon and the remaining constituents are treated as a single spectator particle. Model parameters are determined by fitting the gluon unpolarized and helicity collinear parton distribution functions (PDFs) from global analyses. 
We compare our results with the latest near-threshold $J/\psi$ production data from the GlueX and $J/\psi$-007 experiments at Jefferson Laboratory, finding good agreement for both differential and total cross sections. Predictions are also provided for $\Upsilon$ photoproduction, which can be tested at future Electron-Ion Colliders.

\end{abstract}

\maketitle

\section{Introduction}

Understanding the gluonic structure of hadrons is a central goal in quantum chromodynamics (QCD) and hadronic physics. Gluons, which mediate the strong interaction, dominate the nonperturbative dynamics inside hadrons. However, because they do not couple directly to electroweak probes, gluon distributions are more difficult to access experimentally than those of quarks. Near-threshold exclusive photoproduction of heavy quarkonia, such as $J/\psi$ and $\Upsilon$, has emerged as a particularly promising channel for probing gluon distributions in nucleons~\cite{Voloshin:1978hc,Gottfried:1977gp,Appelquist:1978rt,Bhanot:1979vb,Kharzeev:1995ij,Kharzeev:1998bz,Gryniuk:2016mpk,Hatta:2018ina}. In the heavy-quark limit, contributions from light quarks are strongly suppressed, and the process proceeds predominantly through gluon exchange~\cite{Wang:2022vhr}. Near threshold, the cross section is sensitive to the gluon condensate $\langle P|F^2|P\rangle$ in the nucleon~\cite{Kharzeev:1995ij,Kharzeev:1998bz}, which is closely related to the gluonic trace anomaly contribution to the nucleon mass.

Recent measurements of $\gamma p \rightarrow J/\psi\, p$ near threshold by the GlueX and $J/\psi$-007 collaborations at Jefferson Laboratory (JLab)~\cite{GlueX:2019mkq,GlueX:2023pev,Duran:2022xag} have stimulated considerable theoretical interest. Numerous studies have analyzed this process using different approaches~\cite{Hatta:2019lxo,Mamo:2019mka,Mamo:2022eui,Sun:2021pyw,Sun:2021gmi,Guo:2021ibg,Guo:2023pqw,
Guo:2023qgu,Pentchev:2025qyn,Du:2020bqj,JointPhysicsAnalysisCenter:2023qgg,Sakinah:2024cza,Tang:2024pky,Kim:2025oyo,
Lee:2022ymp,Kim:2024lxc,Mamo:2021krl,Kharzeev:2021qkd}. 
A similar program for $\Upsilon$ photoproduction is planned at future facilities such as the Electron-Ion Collider (EIC) in the U.S.~\cite{Joosten:2018gyo,Accardi:2012qut,Abir:2023fpo,Gryniuk:2020mlh,AbdulKhalek:2021gbh} and the Electron-ion collider in China (EicC)~\cite{Wang:2023thy,Anderle:2021wcy}. In the heavy-quark limit, the photoproduction cross sections can be expressed in terms of the gluon gravitational form factors (GFFs) of the proton~\cite{Pentchev:2025qyn,Mamo:2022eui,Guo:2021ibg,Guo:2023pqw,Guo:2025jiz,Wang:2022ndz}. Together with first-principles lattice QCD calculations of GFFs~\cite{Hackett:2023rif,Pefkou:2021fni}, these measurements offer a unique opportunity to unveil key aspects of the gluonic structure of the proton, including the quantum anomalous energy, mass radius, spin, and mechanical properties~\cite{Ji:1996ek,Polyakov:2018zvc,Ji:2021mtz,Burkert:2023wzr}.

In this work, we adopt the GPD factorization framework for near-threshold heavy quarkonium production~\cite{Guo:2021ibg,Guo:2023pqw,Guo:2023qgu}. Similar to the high-energy regime, the amplitude factorizes into a hard part, the quarkonium distribution amplitude, and gluon GPDs, which reduce to GFFs in the appropriate limit. The large momentum transfer in near-threshold kinematics corresponds to a large skewness parameter $\xi$ for the GPDs. In the limit $\xi \rightarrow 1$, the amplitude is dominated by the leading moments of the GPDs, enabling a systematic asymptotic expansion that relates the cross sections directly to gluon GFFs.

One of the main objectives of hadronic physics is to achieve a three-dimensional picture of hadron structure in terms of quark and gluon degrees of freedom. GPDs provide a unifying framework for this purpose, parameterizing off-forward matrix elements of bilocal QCD operators~\cite{Diehl:2003ny,Belitsky:2005qn}. They encode information on both longitudinal momentum distributions and transverse spatial distributions of partons, as well as their contributions to hadron spin and other properties~\cite{Ji:1996ek,Burkardt:2000za,Bondarenko:2002pp,Burkardt:2002hr,Ralston:2001xs,Burkardt:2005td,Diehl:2002he}. While quark GPDs in nucleons and mesons have been studied extensively~\cite{Pasquini:2005dk,Pasquini:2006dv,Frederico:2009fk,Burkardt:2015qoa,Pasquini:2019evu,Tan:2024doz}, gluon GPDs remain less explored~\cite{Meissner:2007rx,Tan:2023kbl}. Experimentally, GPDs are typically accessed through hard exclusive processes such as deeply virtual Compton scattering~\cite{Ji:1996nm,Collins:1998be,Radyushkin:1997ki} and deeply virtual meson production~\cite{Goeke:2001tz,Ji:1998pc,Collins:1996fb}. In contrast to these high-energy, small-$\xi$ measurements, near-threshold photoproduction of heavy quarkonium offers a unique window into the large-$\xi$ behavior of gluon GPDs.

Here, we compute the differential and total cross sections for near-threshold  photoproduction of $J/\psi$ and $\Upsilon$ by evaluating the relevant gluon GPDs within a spectator model approach~\cite{Lu:2016vqu, Bacchetta:2020vty,Bacchetta:2024fci}. In this model, the proton is treated as a two-body system consisting of an active gluon and a spectator that carries the quantum numbers of the three valence quarks. The nonperturbative nucleon-gluon-spectator vertex is modeled with two dipolar form factors~\cite{Bacchetta:2020vty}. The model parameters are fixed by reproducing the gluon unpolarized and helicity collinear PDFs from global fits at an initial scale $\mu_0 = 2\ \text{GeV}$.

The remaining part of the paper is organized as follows. In Sec.~\ref{Sec:2}, we introduce the GPD factorization formalism for near-threshold photoproduction of heavy quarkonium  and discuss the parameterization of gluon GPDs. In Section~\ref{Sec:3}, we present the calculation of twist-2 gluon GPDs at nonzero skewness within the spectator model. In Sec.~\ref{Sec:4}, we determine the model parameters and compare our predictions for $J/\psi$ production with that data from JLab, and provide predictions for the photoproduction of $\Upsilon$. Finally, Sec.~\ref{Sec:5} summarizes our findings.

\section{GPD framework for the near-threshold photoproduction of heavy quarkonium}\label{Sec:2}

\begin{figure}
	\centering
	\includegraphics[width=0.43\columnwidth]{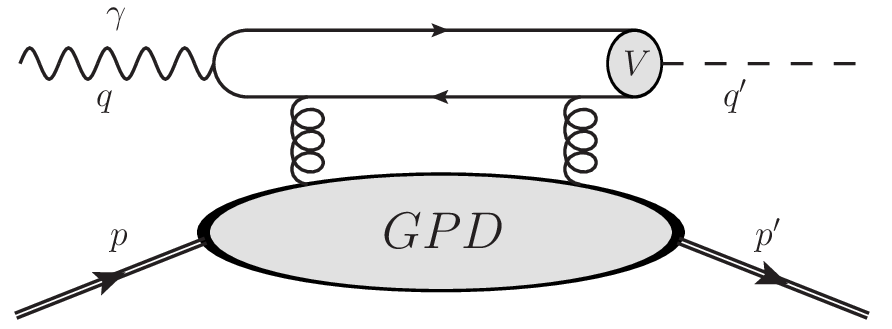}
	\includegraphics[width=0.43\columnwidth]{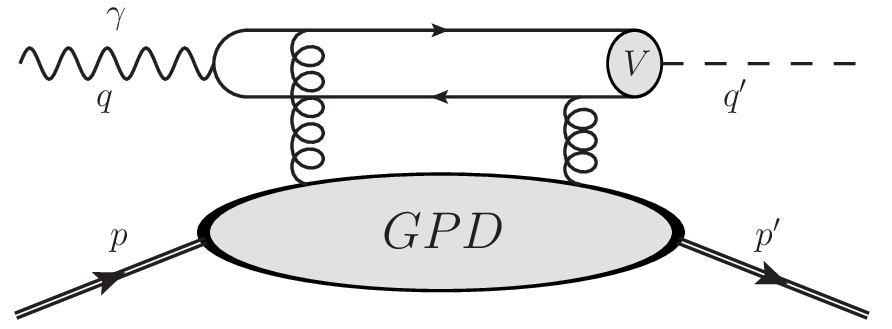}
	\caption{Examples of leading Feynman diagrams that contribute to the exclusive photoproduction of the heavy vector meson.}
	\label{fig:Jpsi}
\end{figure}

\begin{figure}
	\centering
	\includegraphics[width=0.45\columnwidth]{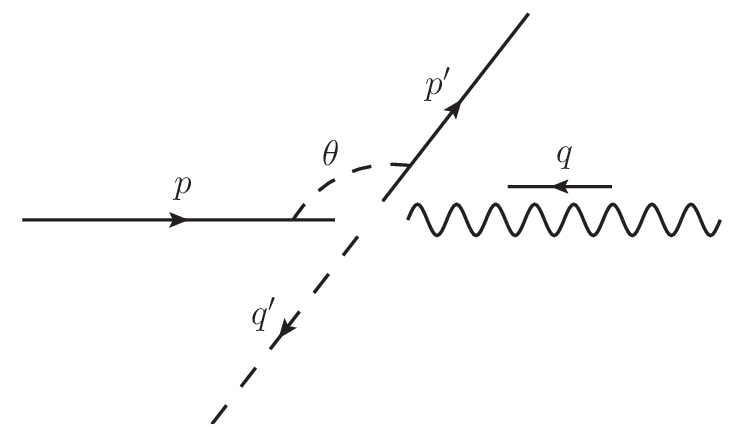}
	\caption{Kinematics for the heavy vector-meson photoproduction in the center-of-mass frame.}
	\label{fig:cm}
\end{figure}

We begin by reviewing the GPD factorization framework for near-threshold photoproduction of heavy quarkonium~\cite{Guo:2021ibg,Guo:2023pqw,Guo:2023qgu}, illustrated in Fig.~\ref{fig:Jpsi}. The kinematics are described by the Mandelstam invariants
\begin{align}
    s = (p+q)^2 \geq (M+M_V)^2, \qquad 
    t = (p^\prime-p)^2 = (q^\prime-q)^2,
\end{align}
where $s = W^2$ is the squared center-of-mass energy, $t = \Delta^2$ is the squared four-momentum transfer, $M$ is the proton mass, and $M_V$ is the produced meson mass. The four-momenta of the incoming photon, incoming proton, outgoing proton, and outgoing meson are denoted by $q$, $p$, $p^\prime$, and $q^\prime$, respectively, satisfying the on-shell conditions:
\begin{align}
	p^{\prime 2}=p^2=M^2,\qquad q^2=-Q^2=0,\qquad q^{\prime 2}=M_V^2.
\end{align}	
Here $Q^2=0$ corresponds to photoproduction; the analysis can be extended to finite virtuality (i.e., large-$Q^2$ leptoproduction) when $Q^2$ is large. We work in the center-of-mass frame of the photon-proton system, as shown in Fig.~\ref{fig:cm}, with results being Lorentz invariant. Choosing the incoming proton along the $+z$ direction, the four-momenta can be expressed in terms of Lorentz scalars as~\cite{He:2024lry,Liu:2024yqa}
\begin{align}
	q=&\bigg(\frac{s-M^2}{2\sqrt{s}},0,0,-\frac{s-M^2}{2\sqrt{s}}\bigg),\nonumber\\
	q^\prime=&\bigg(\frac{s+M_V^2-M^2}{2\sqrt{s}},-|\bm{p}_c^\prime|\sin\theta,0,-|\bm{p}_c^\prime|\cos\theta\bigg),\nonumber\\
	p=&\bigg(\frac{s+M^2}{2\sqrt{s}},0,0,\frac{s-M^2}{2\sqrt{s}}\bigg),\nonumber
\end{align}
\begin{align}	
	p^\prime=\bigg(\frac{s-M_V^2+M^2}{2\sqrt{s}},|\bm{p}_c^\prime|\sin\theta,0,|\bm{p}_c^\prime|\cos\theta\bigg),
\end{align}	
where the magnitude of the outgoing three-momentum $|\bm{p}_c^\prime|$ is expressed as
\begin{align}
	|\bm{p}_c^\prime|=\bigg(\frac{[s-(M_V+M)^2][s-(M_V-M)^2]}{4s}\bigg)^{1/2},
\end{align}	
and the scattering angle $\theta$ is fixed as
\begin{align}
	\cos\theta=\frac{2st+(s-M^2)^2-M_V^2(s+M^2)}{2\sqrt{s}|\bm{p}_c^\prime|(s-M^2)}.
\end{align}	

The skewness parameter, defined as $\xi = -\Delta\cdot q/(2P\cdot q) = (p^+-p^{\prime+})/(p^++p^{\prime+})$ with $P=(p^\prime+p)/2$, evaluates to
\begin{align}
    \xi = \frac{t-M_V^2}{2M^2+M_V^2-2s-t}.
\end{align}	

In the heavy-quark limit $M_V \gg M$ (which implies $W\gg M$), the incoming proton moves nearly along the light-front $+$ direction, defined by $x^{\pm} \equiv (x^0 \pm x^z)/\sqrt{2}$. At threshold, $\sqrt{s} \to M+M_V$, the momentum transfer squared becomes
\begin{align}
	-t_{\text{th}}=\frac{MM_V^2}{M+M_V},
\end{align}	
which is of order $MM_V$ in the heavy-quark limit.

\begin{figure}
	\centering
	\includegraphics[width=0.43\columnwidth]{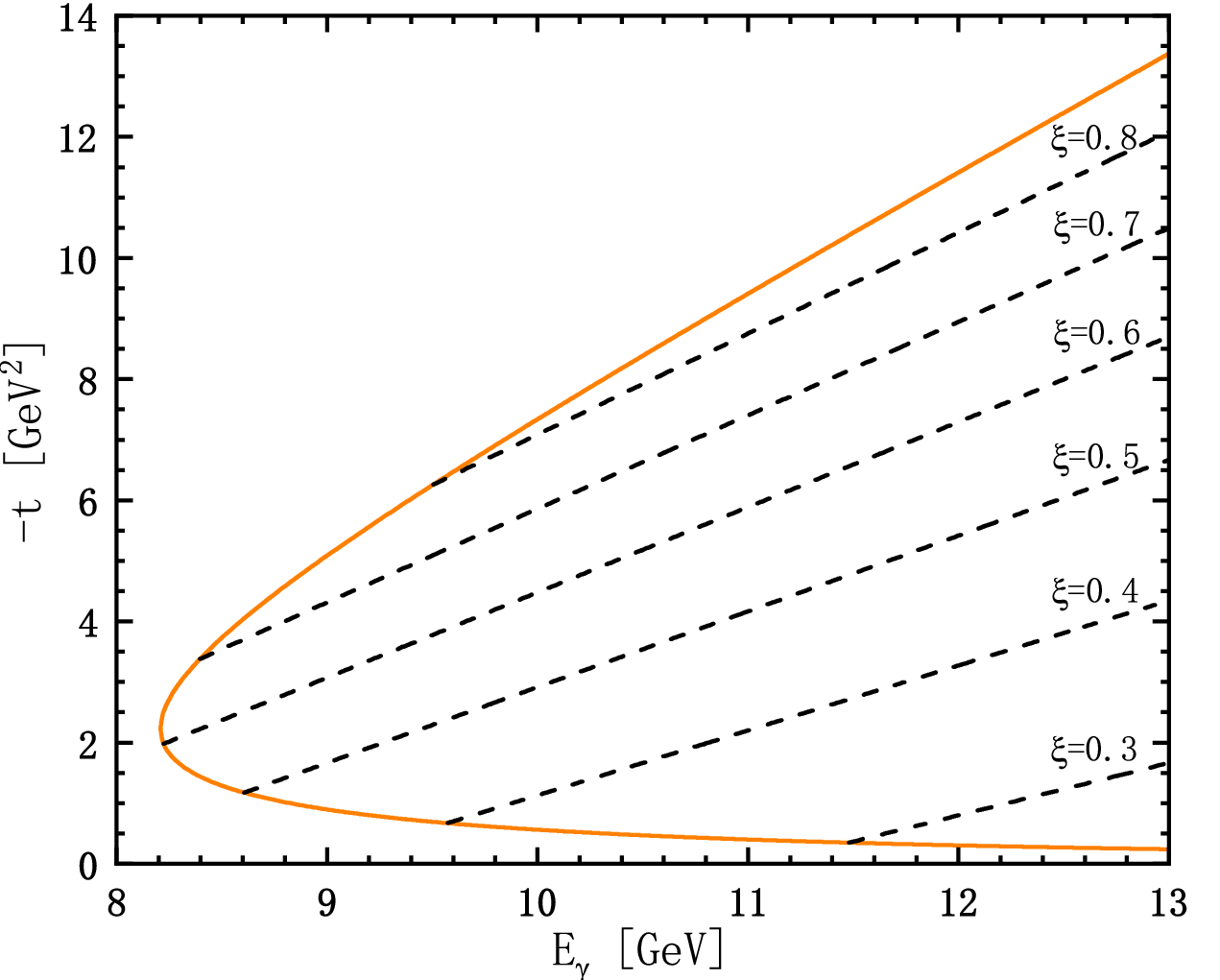}
	\includegraphics[width=0.43\columnwidth]{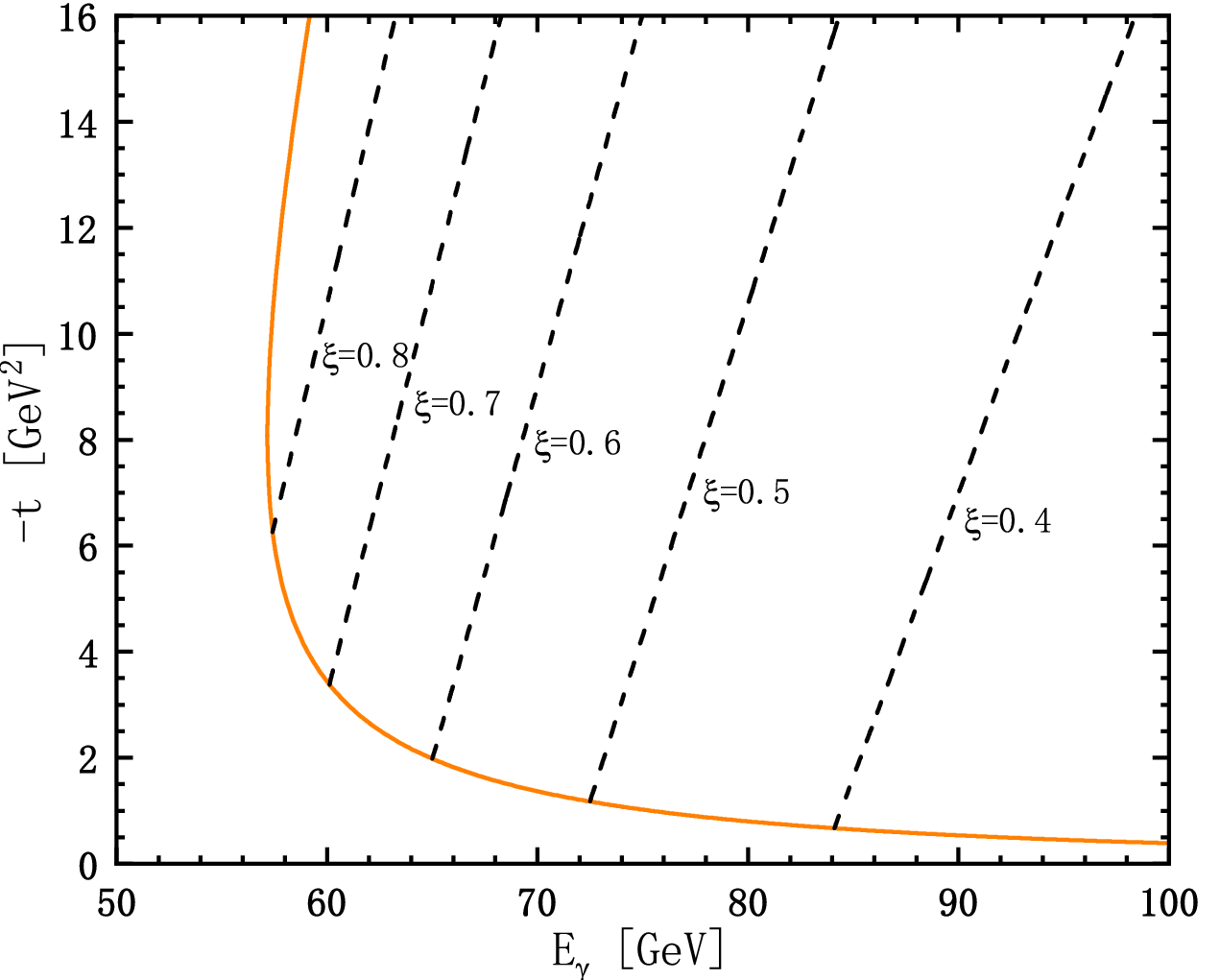}
	\caption{Left panel: $\xi$ values on the $(W,-t)$ plane in the kinematically allowed region with $M_{J/\psi}=3.097\,\text{GeV}$. Right panel: $\xi$ values on the $(W,-t)$ plane in the kinematically allowed region with $M_{\Upsilon}=9.46\,\text{GeV}$.}
	\label{fig:Wt}
\end{figure}

As $W$ increases, the allowed region of $-t$ forms a band bounded by the backward ($\cos\theta=+1$) and forward ($\cos\theta=-1$) limits, denoted as $|t|_{\text{min}}(W)$ and $|t|_{\text{max}}(W)$, respectively. Near threshold, both of them are of order $MM_V$ and significantly larger than $M^2$. 
Figure~\ref{fig:Wt} displays the skewness $\xi$ over the allowed $(W,-t)$ region for $J/\psi$ and $\Upsilon$ production.

A key feature of near-threshold kinematics is that $|t|$ is large and $\xi$ approaches 1 in the heavy-quark limit. Combined with QCD factorization in the GPD framework, this ensures that the photoproduction amplitude of heavy quarkonium is dominated by the lowest moments of the gluon GPDs. 
Therefore, after averaging over initial proton spins and summing over final spins and photon/meson polarizations, the $t$-differential cross section can be written as~\cite{Guo:2021ibg,Guo:2023pqw,Guo:2023qgu}
\begin{align} 
\frac{d\sigma}{dt}=\frac{\alpha_\text{EM}e^2_\text{Q}}{4(W^2-M^2)^2}
\frac{(16\pi\alpha_S)^2}{3M_V^3}|\psi_\text{NR}(0)|^2|G(t,\xi)|^2,
	\label{eq:dsigma}
\end{align}	
where $e_Q$ is the heavy-quark electric charge in units of the proton charge, $\alpha_{\text{EM}}$ and $\alpha_S$ are the electromagnetic and strong running couplings, respectively, and $\psi_{\text{NR}}(0)$ stands for the non-relativistic quarkonium wave function at the origin in the non-relativistic QCD framework~\cite{Bodwin:1994jh}. 
In the unpolarized case, the hadronic matrix element squared  $|G(t,\xi)|^2$ reads
\begin{align}
	|G(t,\xi)|^2=\frac{4}{\xi^4}\bigg\{\bigg(1-\frac{t}{4M^2}\bigg)
    E_2^2-2E_2(H_2+E_2)+(1-\xi^2)(H_2+E_2)^2\bigg\},
	\label{eq:Gt}
\end{align}	
where $H_2\equiv H_2(t,\xi)$ and $E_2\equiv E_2(t,\xi)$ are defined as the leading moments of GPDs,
\begin{align}
	\int_0^1 dx H^g(x,\xi,t)&=A^g(t)+(2\xi)^2C^g(t)\equiv H_2(t,\xi),\nonumber\\
	\int_0^1 dx E^g(x,\xi,t)&=B^g(t)-(2\xi)^2C^g(t)\equiv E_2(t,\xi),
	\label{eq:H2E2}
\end{align}	
with $A^g(t)$, $B^g(t)$, and $C^g(t)$ being the gluon GFFs parameterizing the off-forward matrix element of the gluon energy-momentum tensor~\cite{Polyakov:2018zvc}. Here $x$ denotes the longitudinal momentum fraction of the active gluon. The renormalization scales for the wave function and GPDs have been omitted and should be taken of order $M_V$.

The expressions in Eqs.~(\ref{eq:dsigma}-\ref{eq:Gt}) are evaluated at a generic $\xi$. As mentioned above, near the threshold $\xi$ should be set to 1 in principle. 
however, physical values are $\xi \sim 0.6$ at the $J/\psi$ threshold and $\xi \sim 0.8$ at the $\Upsilon$ threshold. 
The authors of Ref.~\cite{Guo:2021ibg} have confirmed the validity of these expressions for generic $\xi$ by showing that they generalize the results  of Ref.~\cite{Ivanov:2004vd}. Moreover, finite meson mass corrections, particularly important for $J/\psi$, can be partially incorporated by using the physical $\xi$.

The total cross section is obtained by integrating Eq.~\eqref{eq:dsigma} over the kinematically allowed $t$-range: 
\begin{align}
	\sigma=\int^{t_{\text{max}}}_{t_{\text{min}}} dt \left(\frac{d\sigma}{dt}\right).
	\label{eq:sigma}
\end{align}	

The complete set of gluon GPDs for a spin-$1/2$ target was derived in Ref.~\cite{Lorce:2013pza}. The corresponding light-cone correlator is defined as the off-forward matrix element of a bilocal operator~\cite{Diehl:2003ny}:
\begin{align}
	F^{g[ij]}(x,\Delta;\lambda,\lambda^\prime)=\int \frac{dz^-}{2\pi} e^{ik \cdot z} \langle{p^\prime; \lambda^\prime}| F^{+j}_a\left(-\frac{z}{2}\right) \mathcal{W}_{ab}\left(-\frac{z}{2};\frac{z}{2}\right) F^{+i}_b\left(\frac{z}{2}\right)|p; \lambda \rangle \big|_{z^+=0^+,\bm{z}_\perp=\bm{0}_\perp},
	\label{eq:correlator}
\end{align}
where one-particle states are normalized covariantly as $\langle p^\prime|p\rangle=2p^0(2\pi)^3\delta^{(3)}(\bm{p}^\prime-\bm{p})$, $\lambda$ and $\lambda^\prime$ denote the helicities of the incoming and outgoing protons, respectively; $i,j$ are transverse indices, $a,b$ are color indices, and
\begin{align}
	\mathcal{W}_{ab} \left(-\frac{z}{2};\frac{z}{2} \right) \bigg|_{z^+=0^+,\bm{z}_\perp=\bm{0}_\perp}=\left[0^+,-\frac{z^-}{2},\bm{0}_\perp;0^+,\frac{z^-}{2},\bm{0}_\perp\right]_{ab}=\mathcal{P}\exp\left[-g\int^{\frac{z^-}{2}}_{-\frac{z^-}{2}} dy^- f_{abc} A^+_c(0^+,y^-,\bm{0}_\perp)\right]
\end{align}
is the Wilson line that ensures color gauge invariance of the correlator. Here $\mathcal{P}$ denotes path ordering, and $F_{a}^{\mu\nu}$ is the gluon field strength tensor. 
In the light-front gauge ($A^+=0$), the Wilson line reduces to unity.
The twist-2 gluon GPDs for an unpolarized proton appear in the parametrization of the correlator in Eq.~(\ref{eq:correlator}) according to~\cite{Diehl:2003ny}:
\begin{align}
	F^g(x,\Delta;\lambda,\lambda^\prime)=\delta^{ij}_\perp F^{g[ij]}(x,\Delta;\lambda,\lambda^\prime)= \frac{1}{2}\bar{u}(p^\prime,\lambda^\prime)
	\left(\gamma^+H^{g}(x,\xi,t)+\frac{i\sigma^{+\mu}\Delta_\mu}{2M}E^{g}(x,\xi,t)\right)u(p,\lambda),
	\label{eq:Fg}
\end{align}	
where $\delta^{ij}_\perp = -g_\perp^{ij}$ and $\sigma^{\mu\nu} = i[\gamma^\mu,\gamma^\nu]/2$.

\section{Gluon GPDS in the Spectator Model}\label{Sec:3}

In this section, we present an analytical calculation of the gluon GPDs $H^g(x,\xi,t)$ and $E^g(x,\xi,t)$ at nonzero skewness $\xi$ in the proton using a spectator model~\cite{Lu:2016vqu}. A variant of this model, incorporating a spectral function for the spectator mass, has previously been used to compute twist-2  gluon TMD distributions~\cite{Bacchetta:2020vty,Bacchetta:2024fci}.  
For the following discussion, it is convenient to introduce the polarization states of the proton in an arbitrary direction:
\begin{align} 	|p;S\rangle=&\cos\left(\frac{\alpha}{2}\right)|p;+\rangle+\sin\left(\frac{\alpha}{2}\right)e^{i\phi}|p;-\rangle,\nonumber\\
\langle p^\prime;S|=&\cos\left(\frac{\alpha}{2}\right)\langle p^\prime;+|+\sin\left(\frac{\alpha}{2}\right)e^{-i\phi}\langle p^\prime;-|,	
\end{align}
which express  the incoming (outgoing) proton states with both longitudinal and transverse polarizations as a superposition of states with definite light-cone helicities $|p;+\rangle$ and $|p;-\rangle$ ($\langle p^\prime;+|$ and $\langle p^\prime;-|$)~\cite{Diehl:2005jf}. In the rest frame of the proton, the state $|p;S\rangle$ describes a particle whose three-dimensional spin vector is
\begin{align}
	\bm{S}=(S_\perp^1,S_\perp^2,\lambda)=(\sin\alpha \cos\phi,\sin\alpha \sin\phi,\cos\alpha).
\end{align}	

Replacing the helicity states in the matrix element~(\ref{eq:correlator}) according to
\begin{align}
	\langle p^\prime;\lambda^\prime |\rightarrow\langle p^\prime;S|,\qquad |p;\lambda\rangle\rightarrow |p;S\rangle,
\end{align}	
a general correlator $F(x,\Delta;S)$ for arbitrary spin orientations can be constructed by anppropriate choice of the spin vector~\cite{Meissner:2007rx},
\begin{align}	
	F(x,\Delta;S)=&\frac{1}{2}[F(x,\Delta;+,+)+F(x,\Delta;-,-)]
	+\frac{1}{2}\lambda[F(x,\Delta;+,+)-F(x,\Delta;-,-)]\nonumber\\ &+\frac{1}{2}S_\perp^1[F(x,\Delta;-,+)+F(x,\Delta;+,-)]
	+\frac{i}{2}S_\perp^2[F(x,\Delta;-,+)-F(x,\Delta;+,-)].
\end{align}
The explicit calculations employ the standard light-front helicity spinors for the proton~\cite{Lepage:1980fj}:
\begin{align}
	u(p,+)=\frac{1}{\sqrt{2^{3/2}p^+}}\begin{pmatrix}
		&\sqrt{2}p^++M\\
		&p_\perp^1+ip_\perp^2\\
		&\sqrt{2}p^+-M\\
		&p_\perp^1+ip_\perp^2
	\end{pmatrix},~~~~
	u(p,-)=\frac{1}{\sqrt{2^{3/2}p^+}}\begin{pmatrix}
		&-p_\perp^1+ip_\perp^2\\
		&\sqrt{2}p^++M\\
		&p_\perp^1-ip_\perp^2\\
		&-\sqrt{2}p^++M
	\end{pmatrix}.
\end{align}

\begin{figure}
	\centering
	\includegraphics[width=0.45\columnwidth]{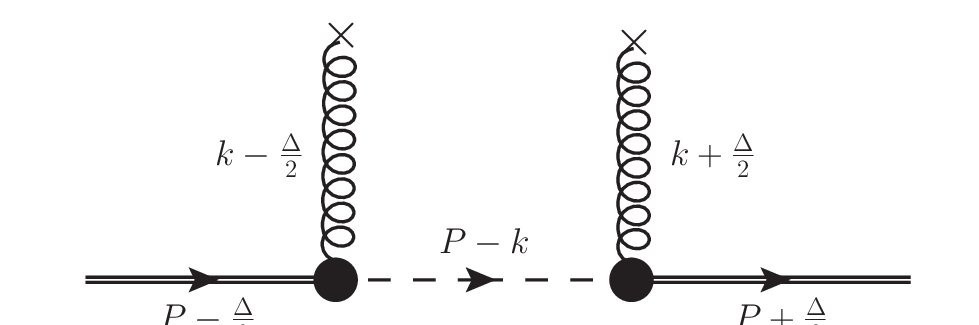}
	\caption{Kinematics for gluon GPDs in the spectator model. The gluon lines with crosses correspond to the specific Feynman rules for the gluon field strength tensor.
}
	\label{fig:GPD}
\end{figure}

The analysis in Ref.~\cite{Wang:2022vhr} revealed that the two-gluon exchange mechanism dominates photoproduction of heavy quarkonium and yields numerical results consistent with the experimental data of near-threshold $J/\psi$ photoproduction.
Fig.~\ref{fig:GPD} shows the Feynman diagram at leading order used to compute the twist-2 gluon GPDs in Eq.~(\ref{eq:Fg}). The proton is modeled as a bound state consisting of an active gluon carrying  momentum $k\pm\frac{\Delta}{2}$ and a single spin-1/2 spectator particle with the momentum $P-k$. 
Therefore, the tree-level scattering amplitude $\mathcal{M}_a^i(S)$ is given by~\cite{Bacchetta:2020vty}
\begin{align}
	\mathcal{M}_a^i(S)&=\langle P-k|F_a^{+i}|p;S\rangle\nonumber\\
	&=\bar{u}_c(P-k)\frac{G_{ab}^{i\mu}(k-\frac{\Delta}{2})}{(k-\frac{\Delta}{2})^2}\mathcal{Y}^-_{\mu,bc}u(p,S),
\end{align}	
where $\bar{u}_c(P-k)$ is the spectator spinor with color index $c$, and
\begin{align} G_{ab}^{i\mu}(k-\frac{\Delta}{2})=-i\delta_{ab}\left(k^+-\frac{\Delta^+}{2}\right)\left(g^{i\mu}-\frac{(k^i-\frac{\Delta^i}{2})(n_-)^\mu}{k^+-\frac{\Delta^+}{2}}\right)
\end{align}	
represents the Feynman rule for the field strength tensor appearing in the definition of the correlator~\cite{Goeke:2006ef,Collins:2011zzd}. 
Here $n_{\pm}$ are the light-like vectors satisfying $n_{\pm}^2=0$ and $n_+\cdot n_-=1$. 
The nucleon-gluon-spectator vertex $\mathcal{Y}_{\mu,bc}^-$ is modeled as~\cite{Bacchetta:2020vty}
\begin{align}
	\mathcal{Y}^-_{\mu,bc}=\delta_{bc}\bigg[g_1^-\gamma_\mu+g_2^-\frac{i}{2M}\sigma_{\mu\nu}p^\nu\bigg],
	\label{eq:vertex}
\end{align}	
where $g_{1,2}^-$ are the model-dependent form factors analogous to Dirac and Pauli form factors, but obviously they are not equivalent because the vertex in Eq.~(\ref{eq:vertex}) involves the nonperturbative color interaction. 
In order to characterize the nonperturbative physics, we adopt the dipolar form factors
\begin{align}
	g_{1,2}^\pm\equiv g_{1,2}\left(\bm{k}_\perp\pm\frac{\bm{\Delta}_\perp}{2}\right)=\kappa_{1,2} \frac{(k\pm\frac{\Delta}{2})^2}{|(k\pm\frac{\Delta}{2})^2-\Lambda_X^2|^2}
=-\kappa_{1,2}\frac{(1-x)[(\bm{k}_\perp\pm\frac{\bm{\Delta}_\perp}{2})^2+L_X^2(0)]}
{[(\bm{k}_\perp\pm\frac{\bm{\Delta}_\perp}{2})^2+L_X^2(\Lambda_X^2)]^2},
\end{align}
where the superscript $+\,(-)$ corresponds to the vertex of the outgoing (incoming) gluon. Here $\kappa_{1,2}$ are the normalization parameters representing the strength of the vertex, $\Lambda_X$ is a cut-off parameter, $\bm{\Delta}_\perp$ represents the transverse component of the momentum transfer, and
\begin{align}
	L_X^2(0)&=xM_X^2-x(1-x)M^2,\nonumber\\
	L_X^2(\Lambda_X^2)&=xM_X^2+(1-x)\Lambda_X^2-x(1-x)M^2,
\end{align}
where $M_X$ is the mass of the spectator, and the gluon mass $M_g$ is set to 0 GeV. 
Additionally, the on‑shell condition for the spectator, $(P-k)^2 = M_X^2$, implies that the gluon is off‑shell:
\begin{align} \bigg(k\pm\frac{\Delta}{2}\bigg)^2&\equiv\tau\bigg(x,\bigg(\bm{k}_\perp\pm\frac{\bm{\Delta}_\perp}{2}\bigg)^2\bigg)
=-\frac{(\bm{k}_\perp\pm\frac{\bm{\Delta}_\perp}{2})^2+L_X^2(0)}{1-x}.
\end{align} 
The introduction of dipolar form factors not only cancels the singularities of gluon propagators, but also ensures the convergence of the correlator after evaluating the $\bm{k}_\perp$ integral.

According to Fig.~\ref{fig:GPD}, the correlator (Eq.~(\ref{eq:Fg})) at the tree level in the spectator model reads
\begin{align}
	F^{g}(x,\Delta;\lambda,\lambda^\prime)=\int \frac{dk^-d^2\bm{k}_\perp}{(2\pi)^4}\frac{\delta_\perp^{ij}\bar{u}(p^\prime,\lambda^\prime)G^{j\nu\ast}_{ab^\prime}(k+\frac{\Delta}{2})\mathcal{Y}^{+\ast}_{\nu,b^\prime c^\prime}i(\slashed{P}-\slashed{k}+M_X)_{cc^\prime}G^{i\mu}_{ab}(k-\frac{\Delta}{2})\mathcal{Y}^-_{\mu,bc}u(p,\lambda)}{D_{\text{GPD}}}
	\label{eq:F},
\end{align}
where
\begin{align} D_{\text{GPD}}=\bigg[\left(k+\frac{\Delta}{2}\right)^2+i\epsilon\bigg]
	\bigg[\left(k-\frac{\Delta}{2}\right)^2+i\epsilon\bigg]
	\bigg[(P-k)^2-M_X^2+i\epsilon\bigg].
\end{align}
Here all color and spinor indices are contracted.

In the calculations, we encounter two types of $k^-$ integrals:
\begin{align}
	I &=\int^{+\infty}_{-\infty}dk^-\frac{1}{D_{\text{GPD}}}=
	\frac{1}{C}\int^{+\infty}_{-\infty}dk^-\frac{1}{(k^--k_1^-)(k^--k_2^-)(k^--k_3^-)},
	\label{eq:I}\\
	I_k&=\int^{+\infty}_{-\infty}dk^-\frac{k^-}{D_{\text{GPD}}}=
	\frac{1}{C}\int^{+\infty}_{-\infty}dk^-\frac{k^-}{(k^--k_1^-)(k^--k_2^-)(k^--k_3^-)}	,
	\label{eq:Ik}
\end{align}
where
\begin{align}
	C=-8(x+\xi)(x-\xi)(1-x)(P^+)^3.
\end{align}
The gluon distributions in $x$ satisfy the Bose symmetry~\cite{Diehl:2003ny,Ji:1998pc}, leading to $H(x,\xi,t)=H(-x,\xi,t)$ and $E(x,\xi,t)=E(-x,\xi,t)$. 
As shown in Eq.~(\ref{eq:H2E2}), the integrals of gluon GPDs over $x$ are restricted to $x\geq 0$, only the two poles at $k_2^-$ and $k_3^-$  in Eqs.~(\ref{eq:I}-\ref{eq:Ik}) need to be considered,
\begin{align}  
	    k_2^-&=\frac{\Delta^-}{2}+\frac{(\bm{k}_\perp
		-\frac{\bm{\Delta}_\perp}{2})^2-i\epsilon}{2(x+\xi)P^+},\\
	k_3^-&=P^--\frac{\bm{k}_\perp^2+M_X^2-i\epsilon}{2(1-x)P^+},
\end{align}
corresponding to the incoming gluon propagator $(k-\Delta/2)$ and the spectator propagator $(P-k)$, respectively. 
The pole positions depend on $x$, so the $k^-$ integration is performed by contour integration in two $x$‑regions, using the replacements
\begin{align}
	\frac{1}{\left(k-\frac{\Delta}{2}\right)^2+i\epsilon}&\rightarrow \frac{-2\pi i}{2(x+\xi)P^+}\delta\left(k^--k_2^-\right),\\
	\frac{1}{(P-k)^2-M_X^2+i\epsilon}&\rightarrow \frac{-2\pi i}{2(1-x)P^+}\delta\left(k^--k_3^-\right).
\end{align}
The integrals then become
\begin{align}
	I&= 
	\begin{cases}
		-\frac{2\pi i}{C}\frac{1}{(k_2^--k_1^-)(k_2^--k_3^-)}, &  0 \leq  x\leq \xi,\\
		\frac{2\pi i}{C}\frac{1}{(k_3^--k_1^-)(k_3^--k_2^-)}, &  \xi< x \leq 1, 
	\end{cases}\\
    I_k &= 
    \begin{cases}
	    -\frac{2\pi i}{C}\frac{k_2^-}{(k_2^--k_1^-)(k_2^--k_3^-)}, &  0 \leq  x\leq \xi,\\
    	\frac{2\pi i}{C}\frac{k_3^-}{(k_3^--k_1^-)(k_3^--k_2^-)}, &  \xi< x \leq 1, 
    \end{cases}
\end{align}
where the region $0\le x\le\xi$ corresponds to a cut through the incoming gluon, and $\xi<x\le1$ to a cut through the spectator.

After selecting the appropriate helicity combinations and carrying out the $k^-$ integrals in Eq.~(\ref{eq:F}), the twist-2 gluon GPDs can be constructed as
\begin{align}
	H(x,\xi,t) &= 
	\begin{cases}
		\frac{1}{64 M^2 \pi^3 \bm{\Delta}_\perp^2}\int d^2\bm{k}_\perp \frac{N_H^{0 \leq  x\leq \xi}}{D_\text{SM}^{0 \leq  x\leq \xi}}, &  0 \leq  x\leq \xi,\\
		\frac{1}{16 M^2 \pi^3 \bm{\Delta}_\perp^2}\int d^2\bm{k}_\perp \frac{N_H^{\xi< x \leq 1}}{D_\text{SM}^{\xi< x \leq 1}}, &  \xi< x \leq 1, 
	\end{cases}
	\label{eq:GPDH}\\
	E(x,\xi,t) &= 
	\begin{cases}
		\frac{1-\xi^2}{8 M \pi^3 \bm{\Delta}_\perp^2}\int d^2\bm{k}_\perp \frac{N_E^{0 \leq  x\leq \xi}}{D_\text{SM}^{0 \leq  x\leq \xi}}, &  0 \leq  x\leq \xi,\\
		\frac{1-\xi^2}{4 M \pi^3 \bm{\Delta}_\perp^2} \int d^2\bm{k}_\perp \frac{N_E^{\xi< x \leq 1}}{D_\text{SM}^{\xi< x \leq 1}}, &  \xi< x \leq 1, 
	\end{cases}
	\label{eq:GPDE}
\end{align}
with $\bm{\Delta}_\perp^2 = -(1-\xi^2)t - 4\xi^2 M^2$. The explicit forms of the numerators $N_H$, $N_E$ and denominators $D_\text{SM}$ are lengthy and are collected in the Appendix.

In the forward limit ($\Delta\to0$), the model resulting gluon unpolarized and helicity PDFs have been derived in Ref.~\cite{Bacchetta:2020vty},
\begin{align}
	f_1^g(x)=&\int d^2\bm{k}_\perp \frac{1} {(2\pi)^3 4xM^2(L_X^2(0) + \bm{k}_\perp^2)^2}\{[2M x g_1 - x(M + M_X) g_2]^2 [(M_X - M(1 - x))^2 + \bm{k}_\perp^2] \nonumber\\
	&+ 2\bm{k}_\perp^2(\bm{k}_\perp^2 + xM_X^2)g_2^2 + 2\bm{k}_\perp^2M^2(1 - x)(4g_1^2 - xg_2^2)\},
	\label{eq:f1g}\\
	g_{1L}^g(x)=&\int d^2\bm{k}_\perp \frac{1}{(2\pi)^3 4M^2(L_X^2(0) + \bm{k}_\perp^2)^2}\{\bm{k}_\perp^2(2 - x)[2M g_1 - (M_X - M)g_2]+ 2M x[M_X - M(1 - x)]^2g_1\nonumber\\
	& - [(M_X + M)x(L_X^2(0) + (1 - x)(M_X - M)^2) + 2M x\bm{k}_\perp^2]g_2\}[2M g_1 - (M_X + M)g_2].
	\label{eq:g1g}
\end{align}	

\section{Numerical results}\label{Sec:4}

\subsection{Parameter determination and derived quantities}

\begin{figure}
	\centering
	\includegraphics[width=0.43\columnwidth]{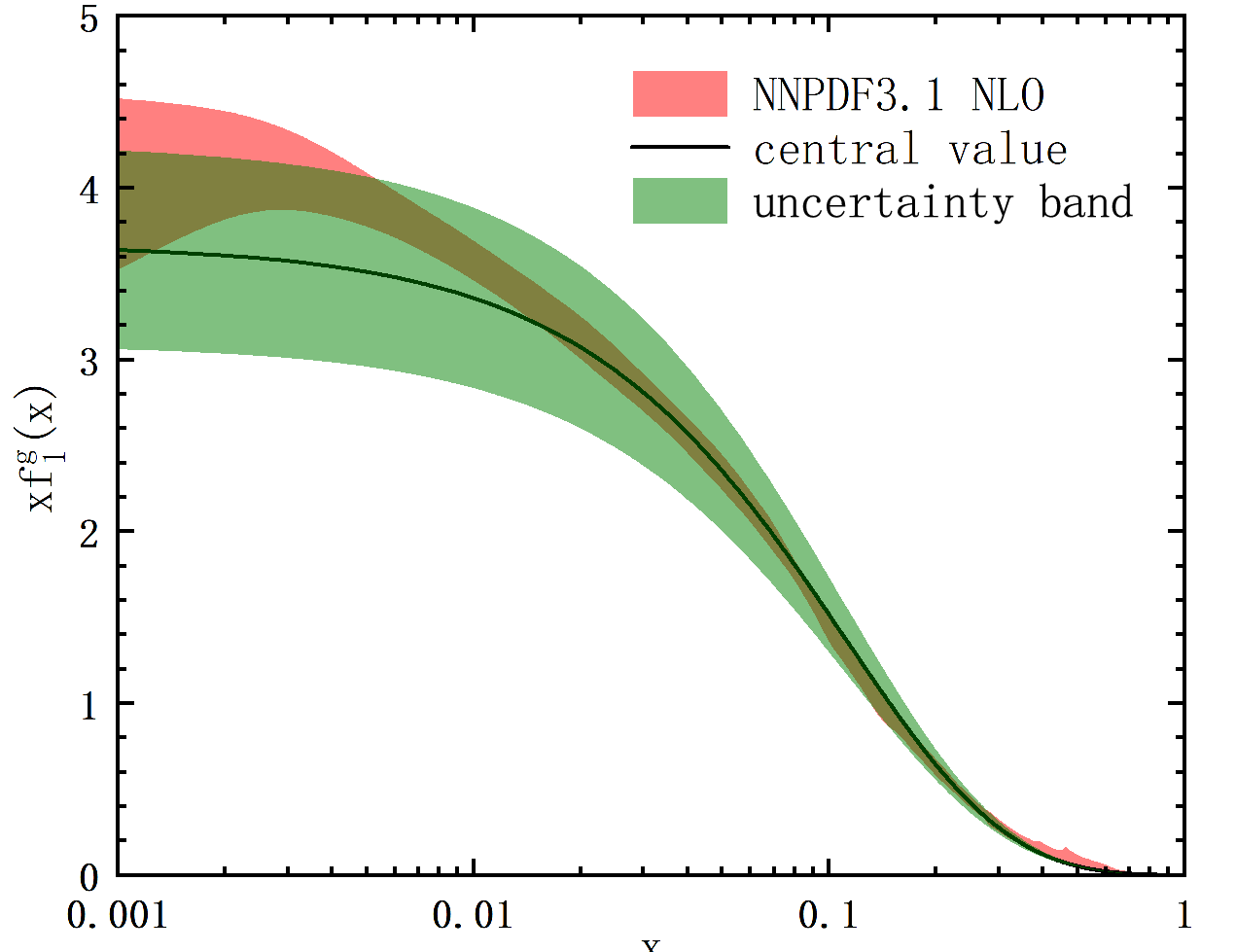}
	\includegraphics[width=0.43\columnwidth]{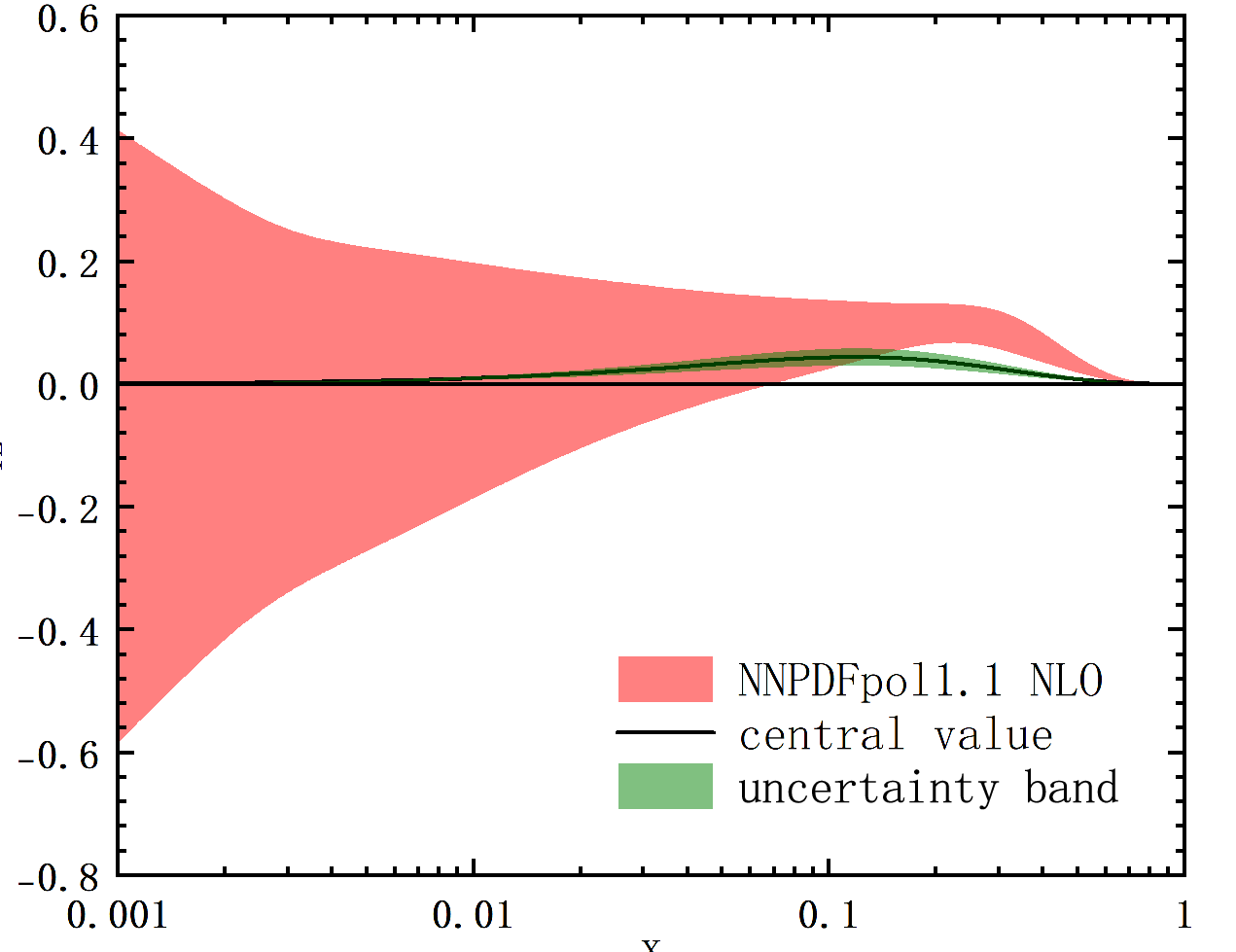}
	\caption{The simultaneous fit of the model results in Eqs.~(\ref{eq:f1g}-\ref{eq:g1g}) to the NNPDF3.1 NLO dataset for $xf_1^g(x)$~\cite{NNPDF:2017mvq} (left panel) and the NNPDFpol1.1 NLO dataset for $xg_{1L}^g(x)$~\cite{Nocera:2014gqa} (right panel).}
	\label{fig:fit}
\end{figure}

The present spectator model adopted in the calculation is similar to that in Ref.~\cite{Bacchetta:2020vty} except we refrain to include a spectral function for the spectator mass. The model contains four free parameters: the normalization constants $\kappa_{1,2}$; the cut‑off $\Lambda_X$ which regulates the large $\bm{k}_\perp$-behavior; and the spectator mass $M_X$ which is taken larger than the proton mass ($M_X>M$).  
To describe the $x$-dependence of the distributions reliably, we determine these parameters by simultaneously fitting the unpolarized PDF (Eq.~\ref{eq:f1g}) and the helicity PDF (Eq.~\ref{eq:g1g}) to the next-to-leading-order (NLO) NNPDF3.1 parametrization of $f_1^g(x)$~\cite{NNPDF:2017mvq} and the NLO NNPDFpol1.1 parametrization of $g_{1L}^g(x)$~\cite{Nocera:2014gqa} at the scale $\mu_0=2\;\text{GeV}$. The fitted results and their uncertainties are listed in Table~\ref{tab1}.

\begin{table}[htbp]
	\centering
	\caption{Model parameters with uncertainties obtained from the simultaneous fit.}
	\label{tab1}
	\begin{tabular}{lcc} 
		\hline   
		Parameter & Central value & Uncertainty \\ 
		\hline 
		~~~~~$\kappa_1$ & 6.503  & $\pm0.388$    \\
		~~~~~$\kappa_2$ & 3.591  & $\pm0.199$    \\
		~~~~~$M_X$ & 1.341  & $\pm0.028$   \\
		~~~~~$\Lambda_X$ & 0.474 & $\pm0.014$   \\
		\hline 
	\end{tabular}
\end{table} 
In Fig.~\ref{fig:fit}, we  the numerical results of the simultaneous fit of $xf_1^g(x)$ (left panel) and $xg_{1L}^g(x)$ (right panel) at $\mu_0=2\,\text{GeV}$. The red bands in the two panels  represent the NNPDF3.1 NLO dataset for $xf_1^g(x)$~\cite{NNPDF:2017mvq} and the NNPDFpol1.1 NLO dataset for $xg_{1L}^g(x)$~\cite{Nocera:2014gqa}, the green bands represent the uncertainty bands of the fit, and the black solid curves represent the results of the central parameter values.

Using the fitted parameters, we first compute the average longitudinal momentum fraction carried by gluons, defined as the second Mellin moment of $f_1^g(x)$ 
\begin{align}
	\langle x\rangle_g=\int^1_0dx xf_1^g(x)=0.421\pm0.060.
	\label{eq:xg}
\end{align}
The result is consistent with the recent lattice result $0.427(92)$ at the same scale~\cite{Alexandrou:2020sml}. Note that 
the GFF $A^g(0)$ is traditionally equivalent to $\langle x\rangle_g$, which is obtained by setting $\xi=0$ and $t=0$ in $H_2(t,\xi)$ (Eq.~\ref{eq:H2E2}).  

Secondly,  the gluon helicity contribution to the proton spin is
\begin{align}
	S_z^g=\int^1_0dx g_{1L}^g(x)=0.121\pm0.038,
\end{align}
and the total gluon angular momentum in the Ji decomposition~\cite{Ji:1996ek} is calculated as
\begin{align}
	J_z^g=\frac{1}{2}\int^1_0dx x[H^g(x,0,0)+E^g(x,0,0)]=0.090\pm0.015,
\end{align}
which is half of the recent lattice value $0.187(46)$~\cite{Alexandrou:2020sml}.

Finally, from the trace of the full energy-momentum tensor~\cite{Polyakov:2018zvc}:
\begin{align}
	H_a&=\langle P|T^\mu\,_\mu|P\rangle\nonumber\\
	   &=\frac{1}{4}(A^q(0)+A^g(0))M,
\end{align}
where $A^q(0)$ is interpreted as the average longitudinal momentum fraction of quarks. we obtain the gluonic contribution to the quantum anomalous energy,
\begin{align}
	H_a^g=\frac{1}{4}A^g(0)M=0.099\pm0.014.
\end{align}	
The good agreement between our model and lattice results for $A^g(0)$ supports the reliability of this estimate, which can be compared with earlier vector‑dominance model analyses~\cite{Kharzeev:1995ij,Wang:2019mza}.


\begin{figure}
	\centering
	\includegraphics[width=0.3\columnwidth]{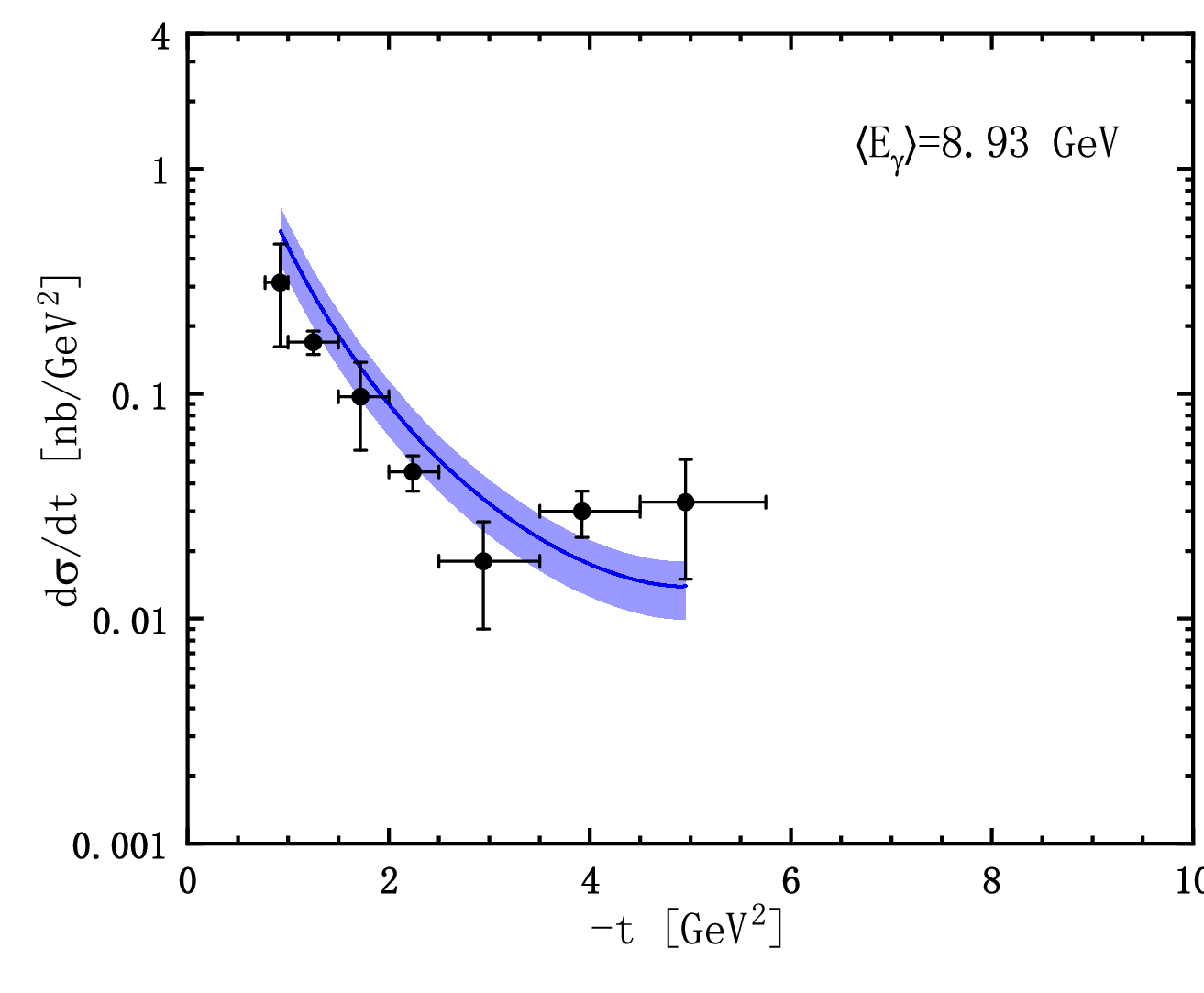}
	\includegraphics[width=0.3\columnwidth]{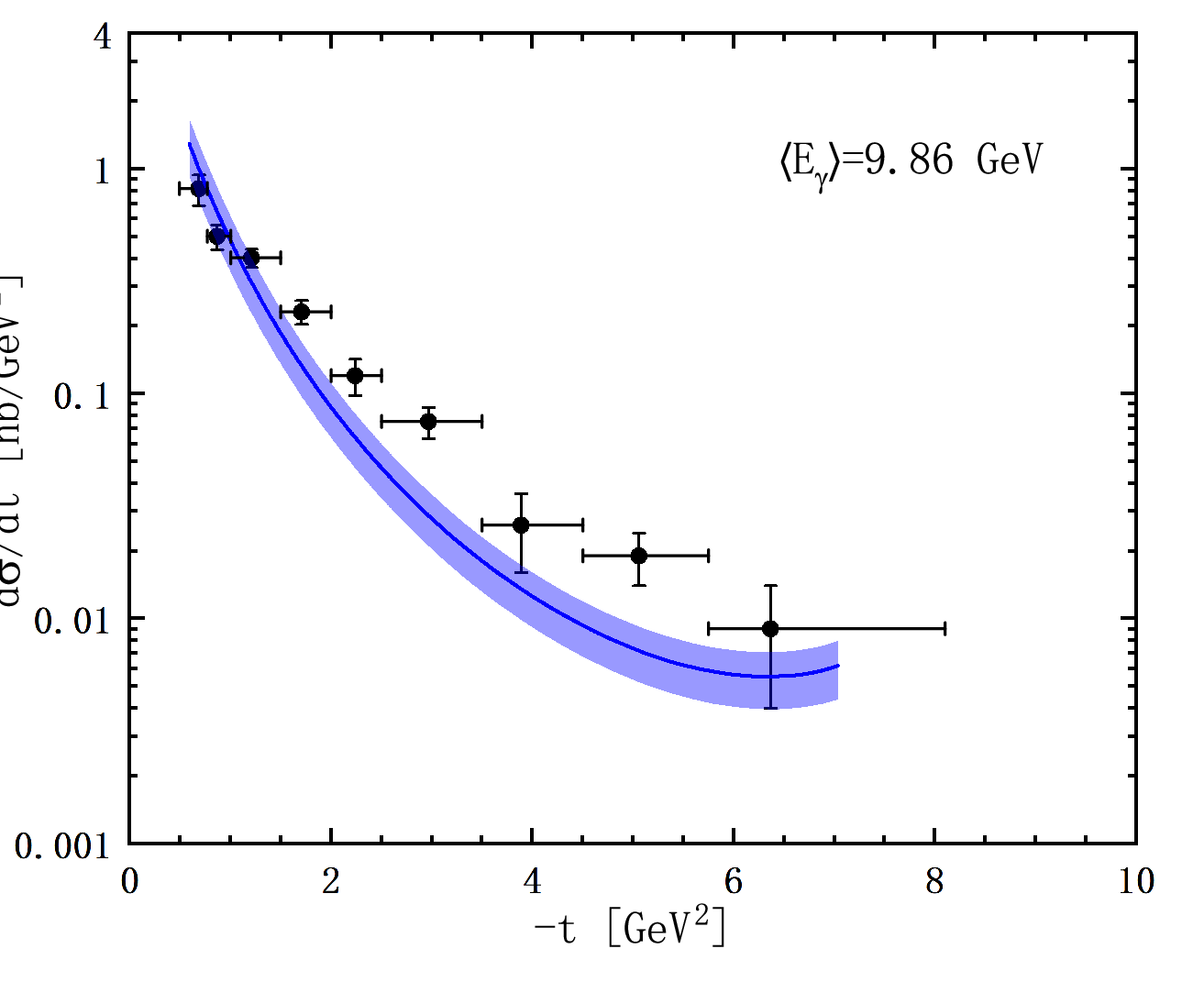}
	\includegraphics[width=0.3\columnwidth]{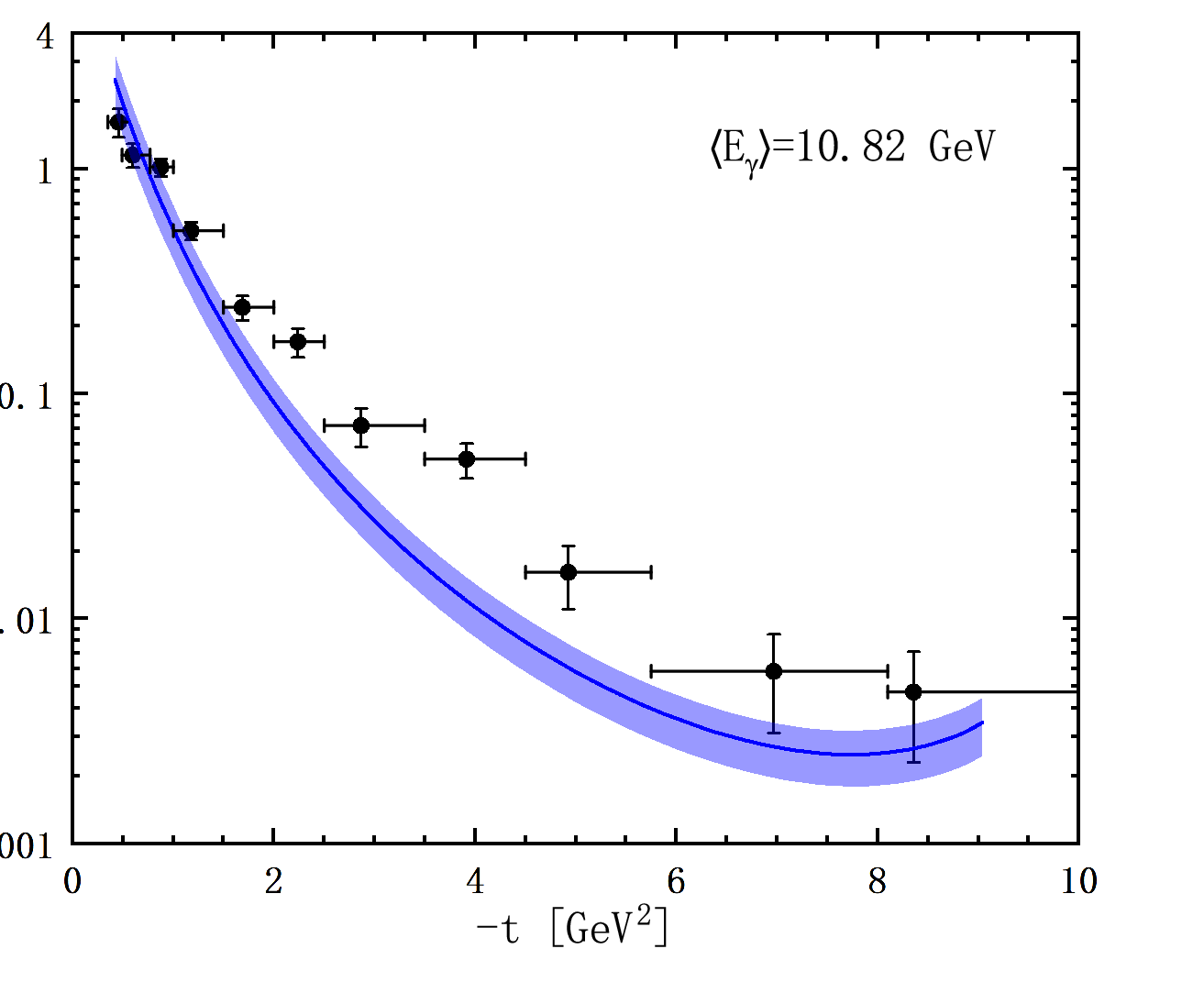}\\
	\includegraphics[width=0.3\columnwidth]{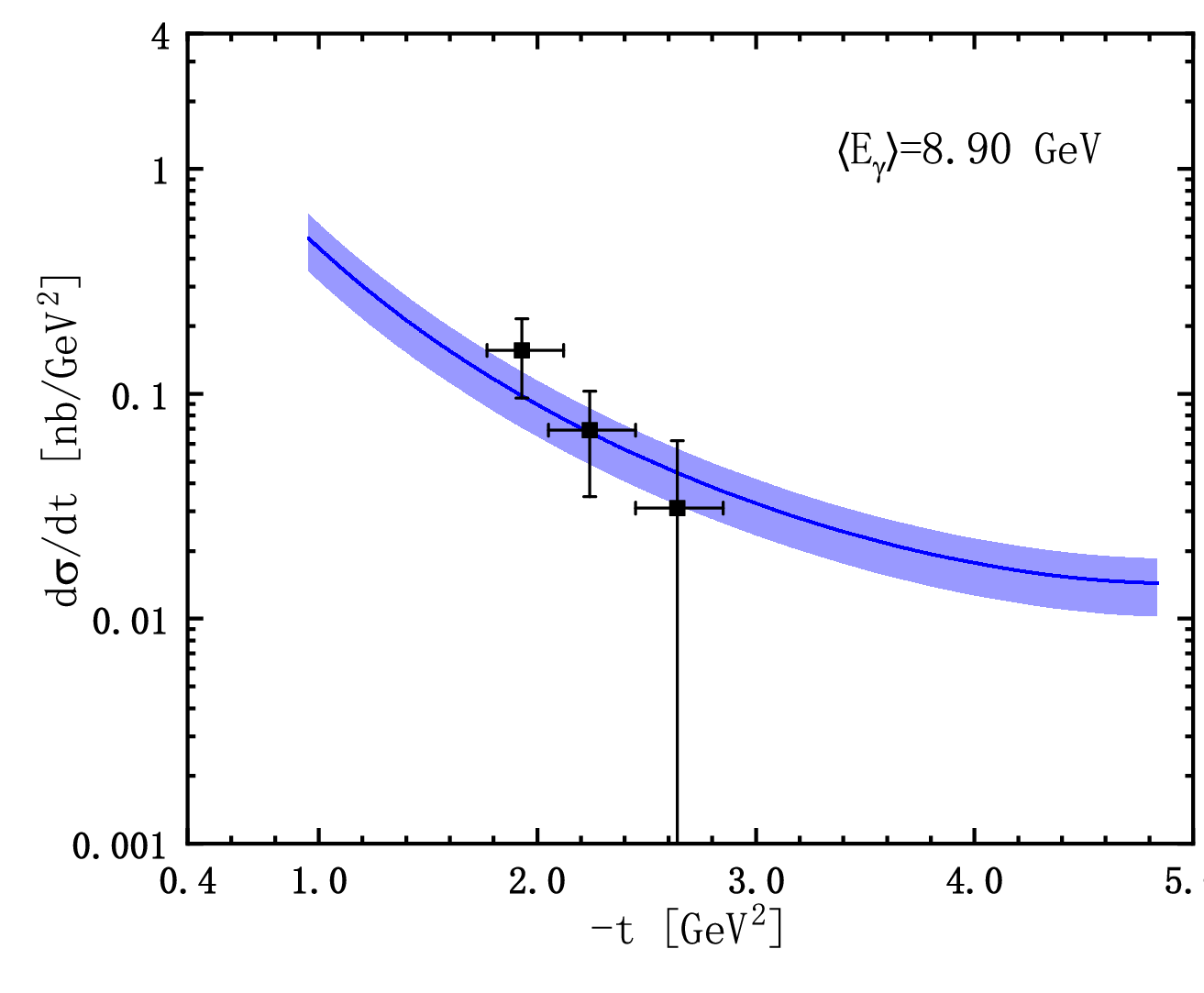} 
	\includegraphics[width=0.3\columnwidth]{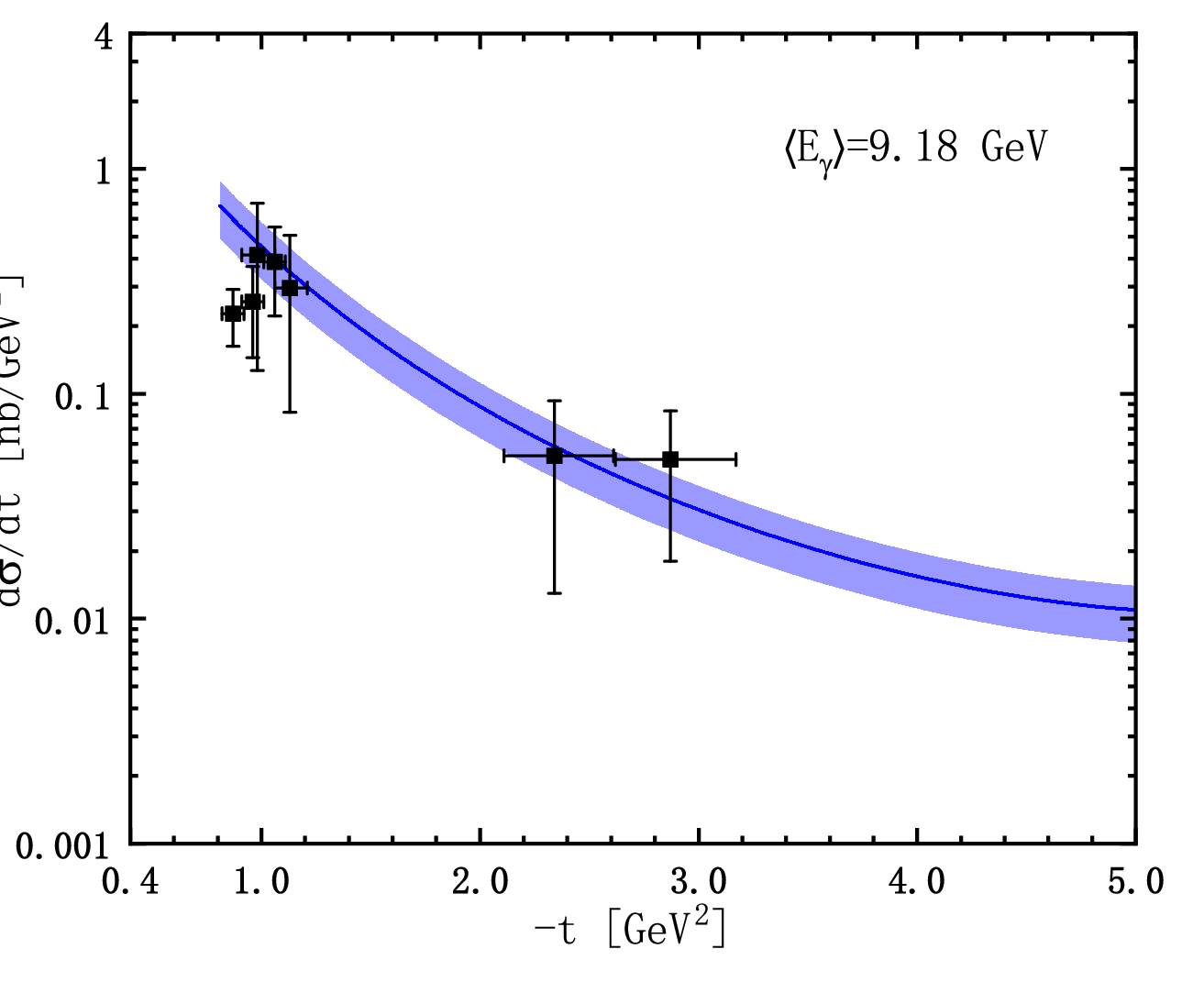}
	\includegraphics[width=0.3\columnwidth]{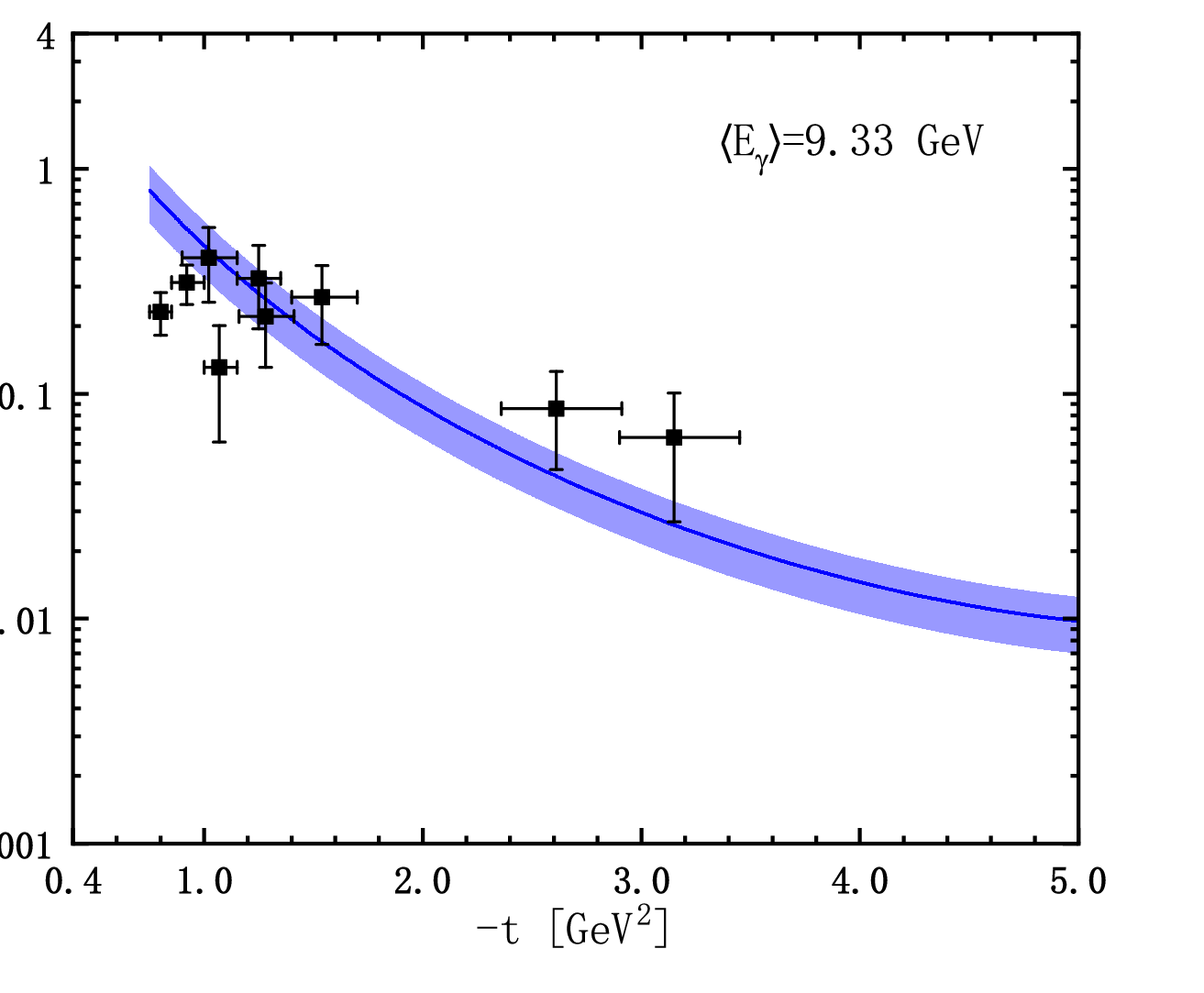}\\
	\includegraphics[width=0.3\columnwidth]{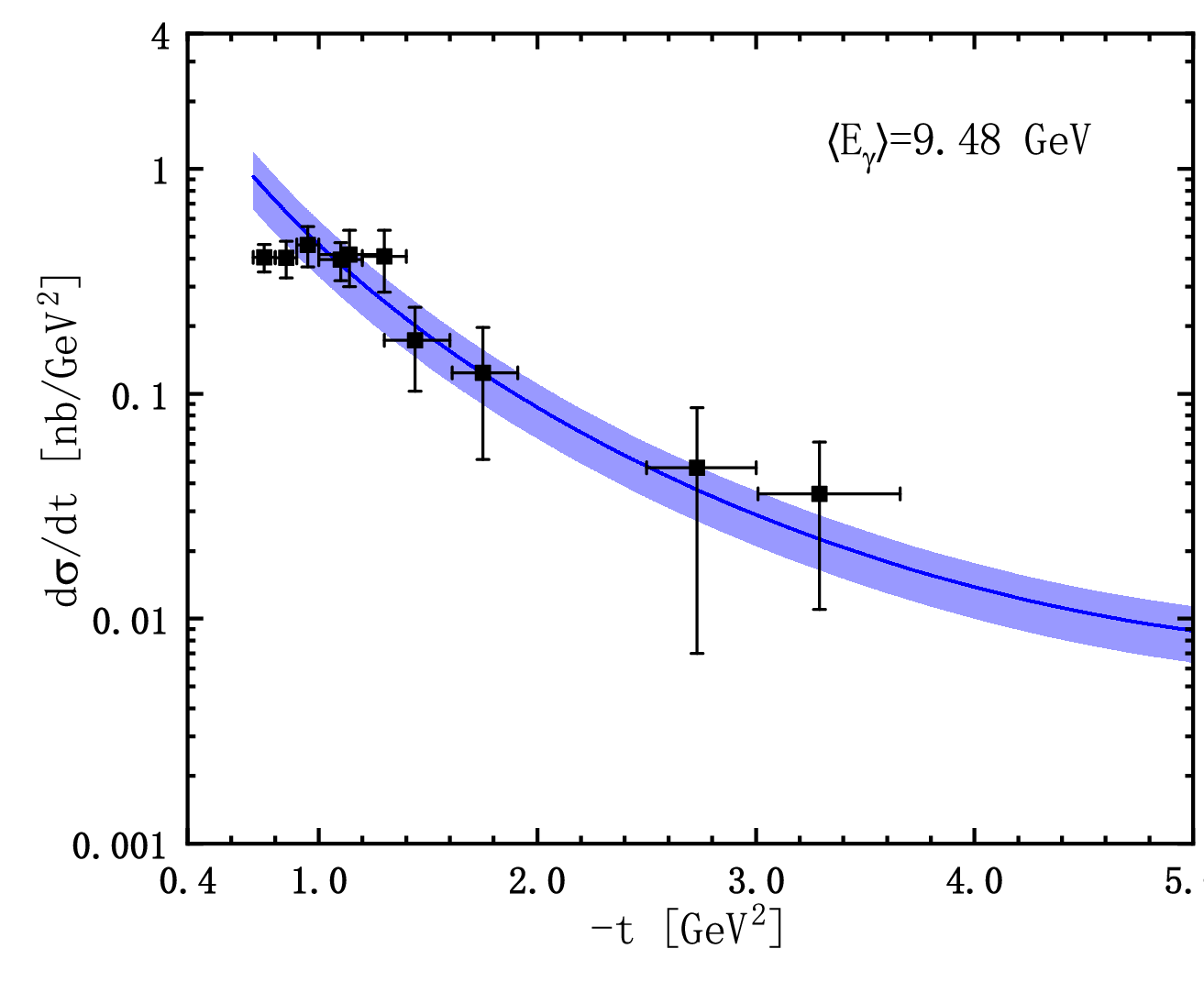}
	\includegraphics[width=0.3\columnwidth]{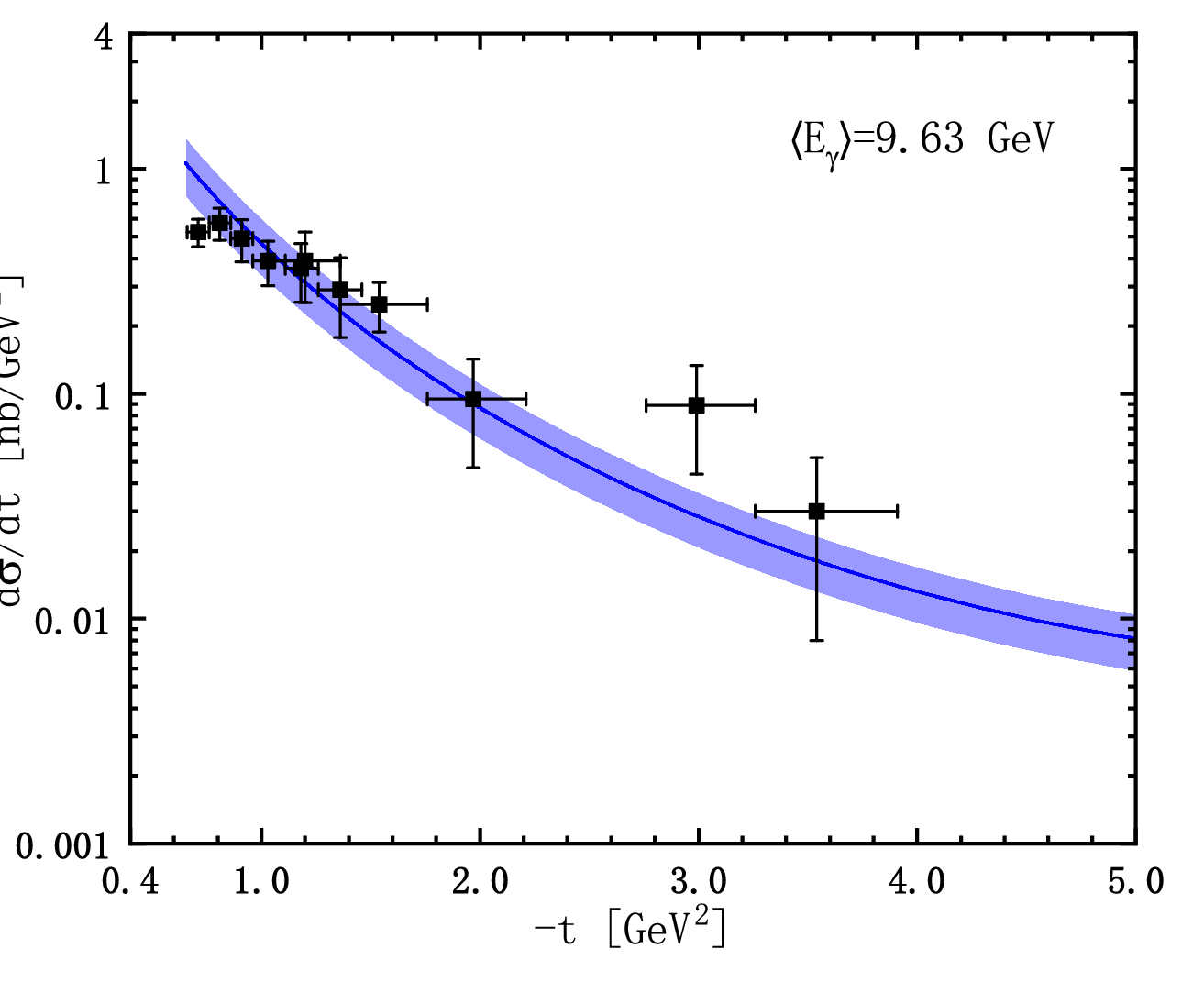}
	\includegraphics[width=0.3\columnwidth]{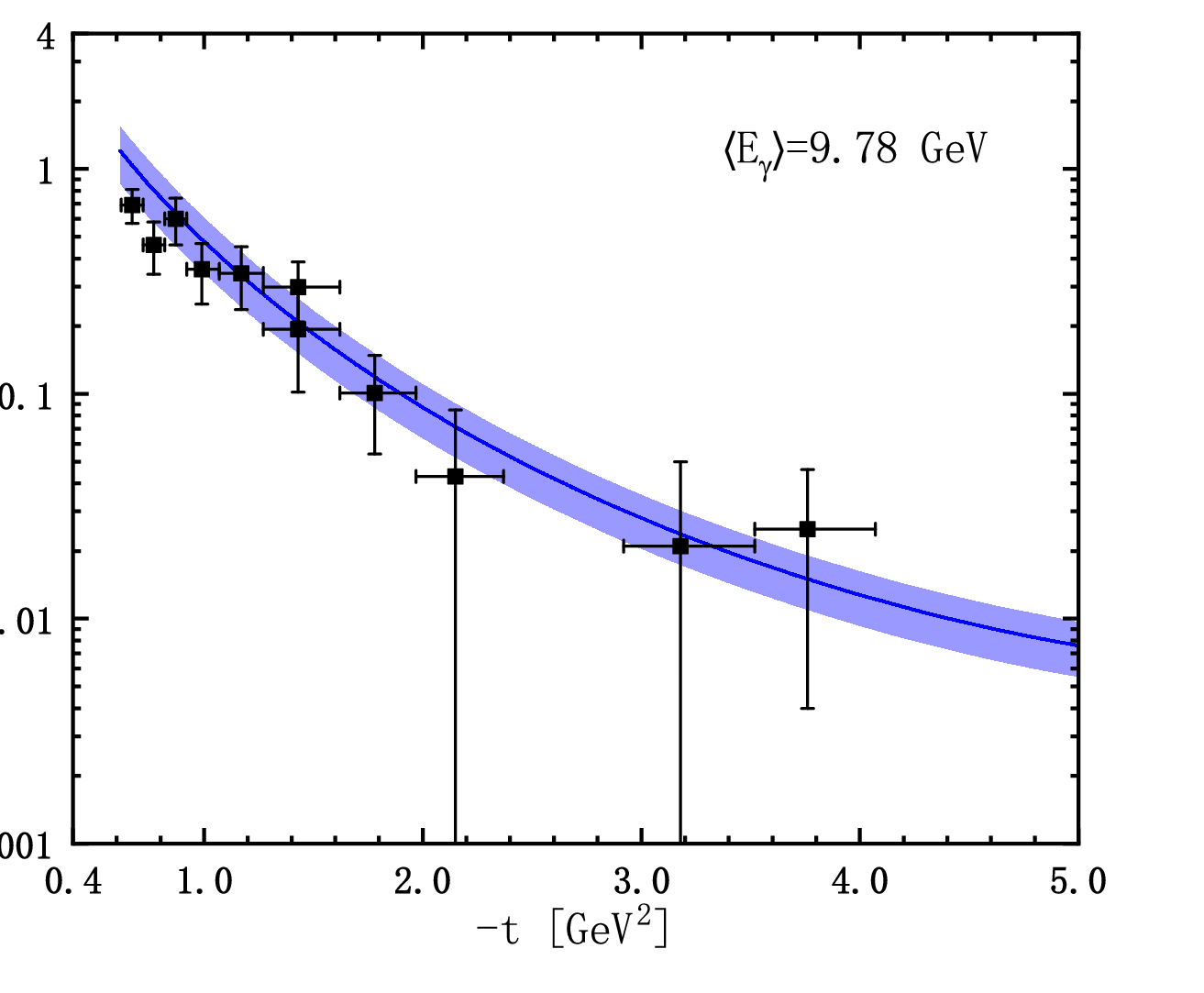}\\
	\includegraphics[width=0.3\columnwidth]{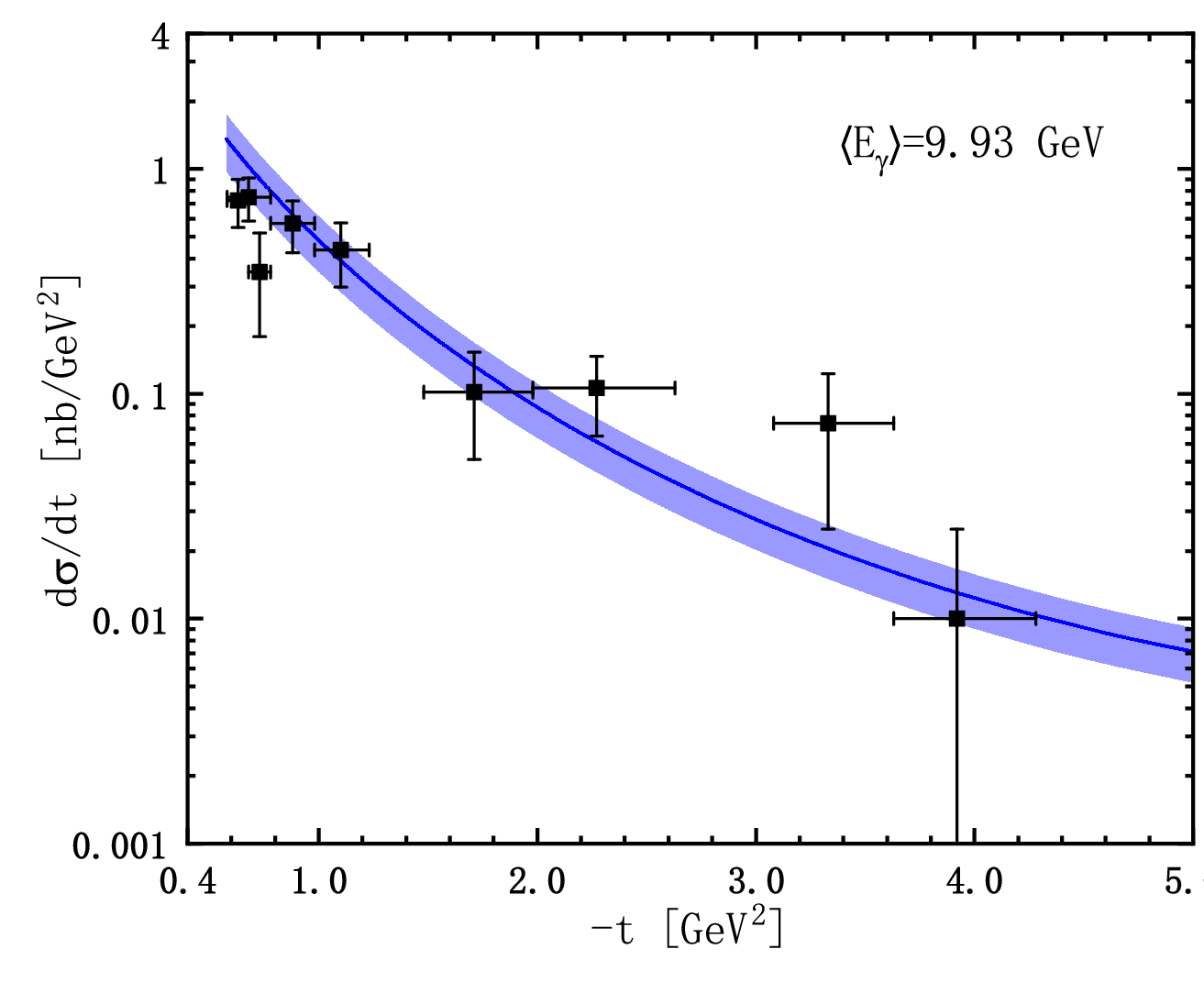}
	\includegraphics[width=0.3\columnwidth]{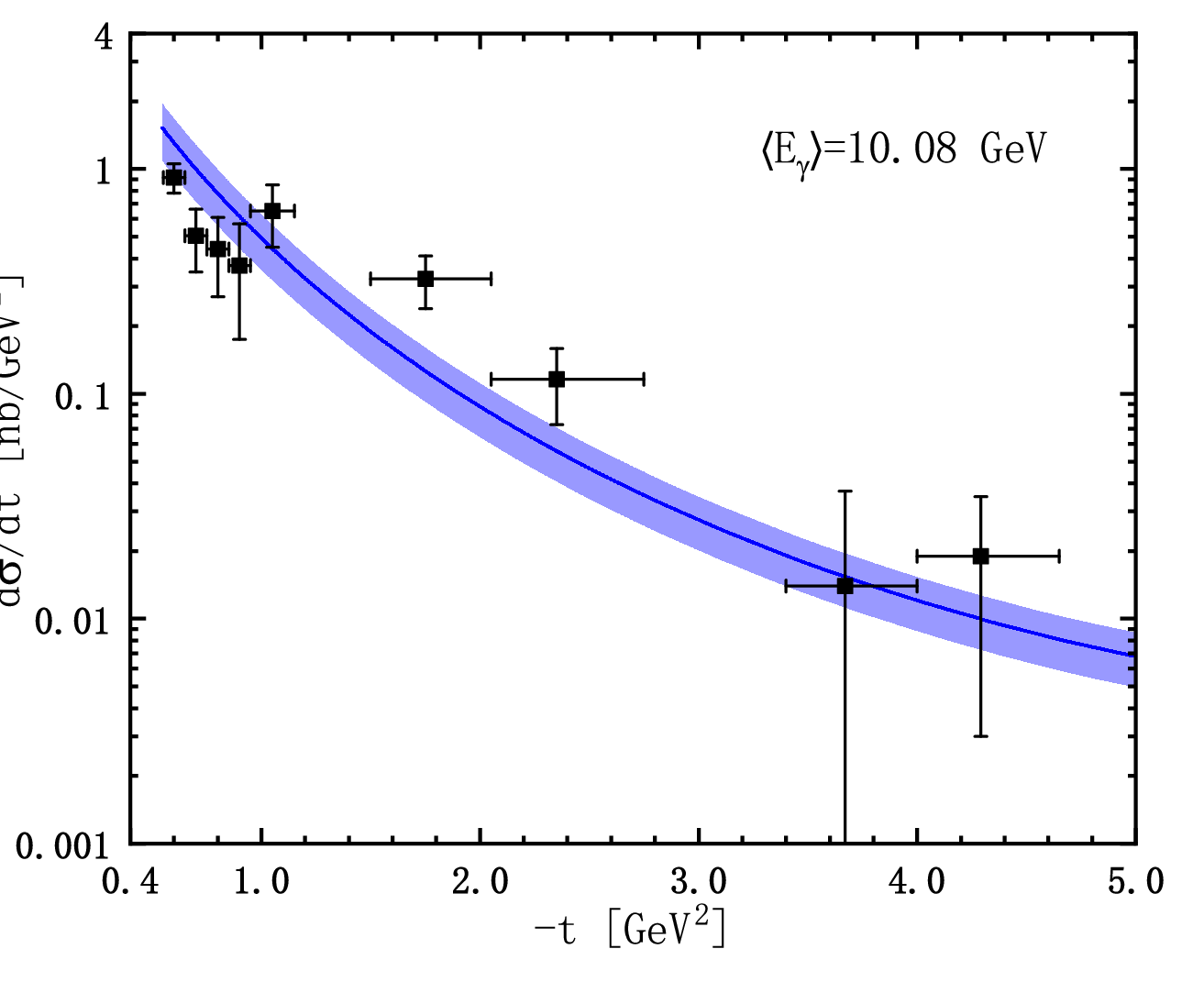}
	\includegraphics[width=0.3\columnwidth]{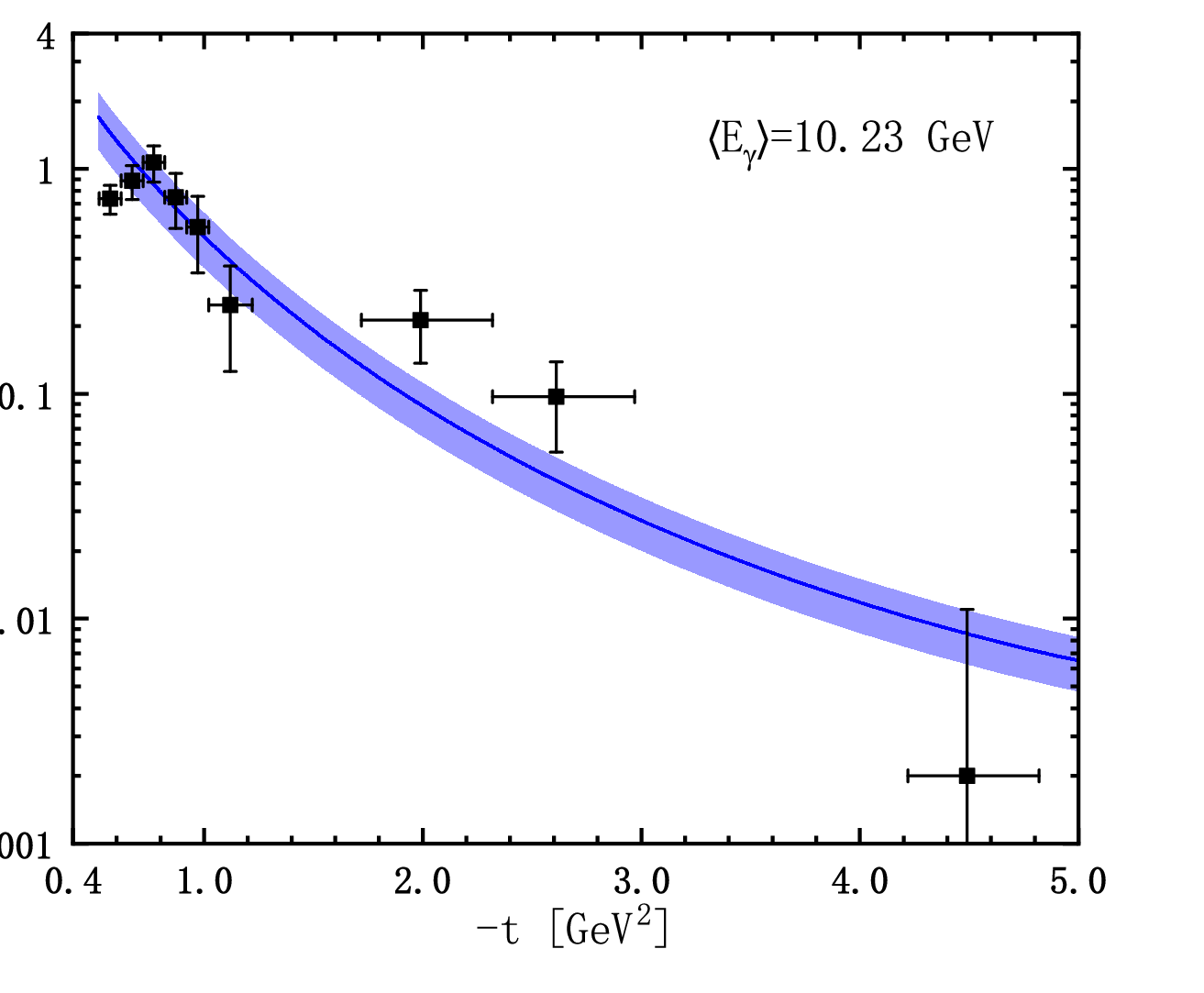}\\
	\includegraphics[width=0.3\columnwidth]{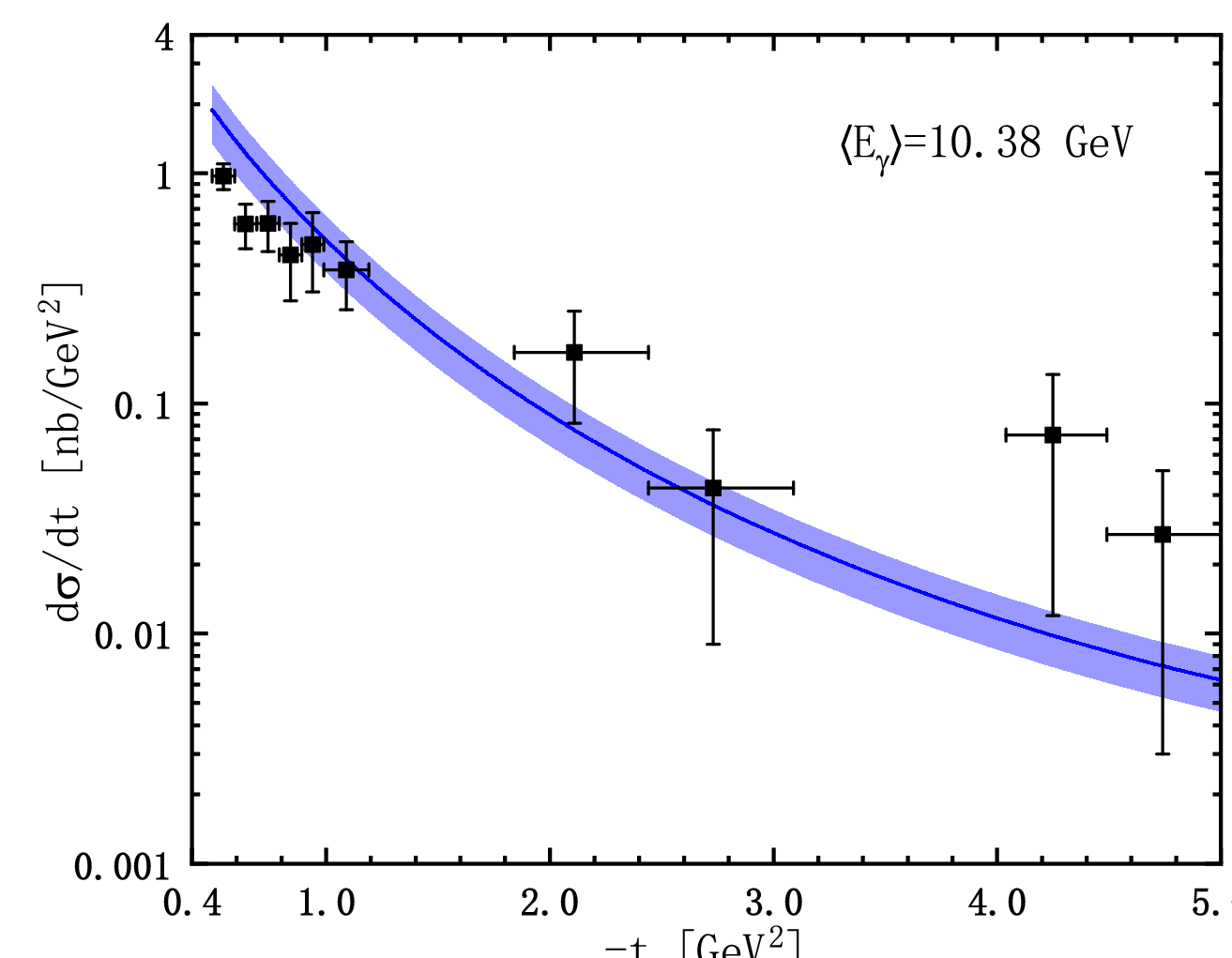}
	\includegraphics[width=0.3\columnwidth]{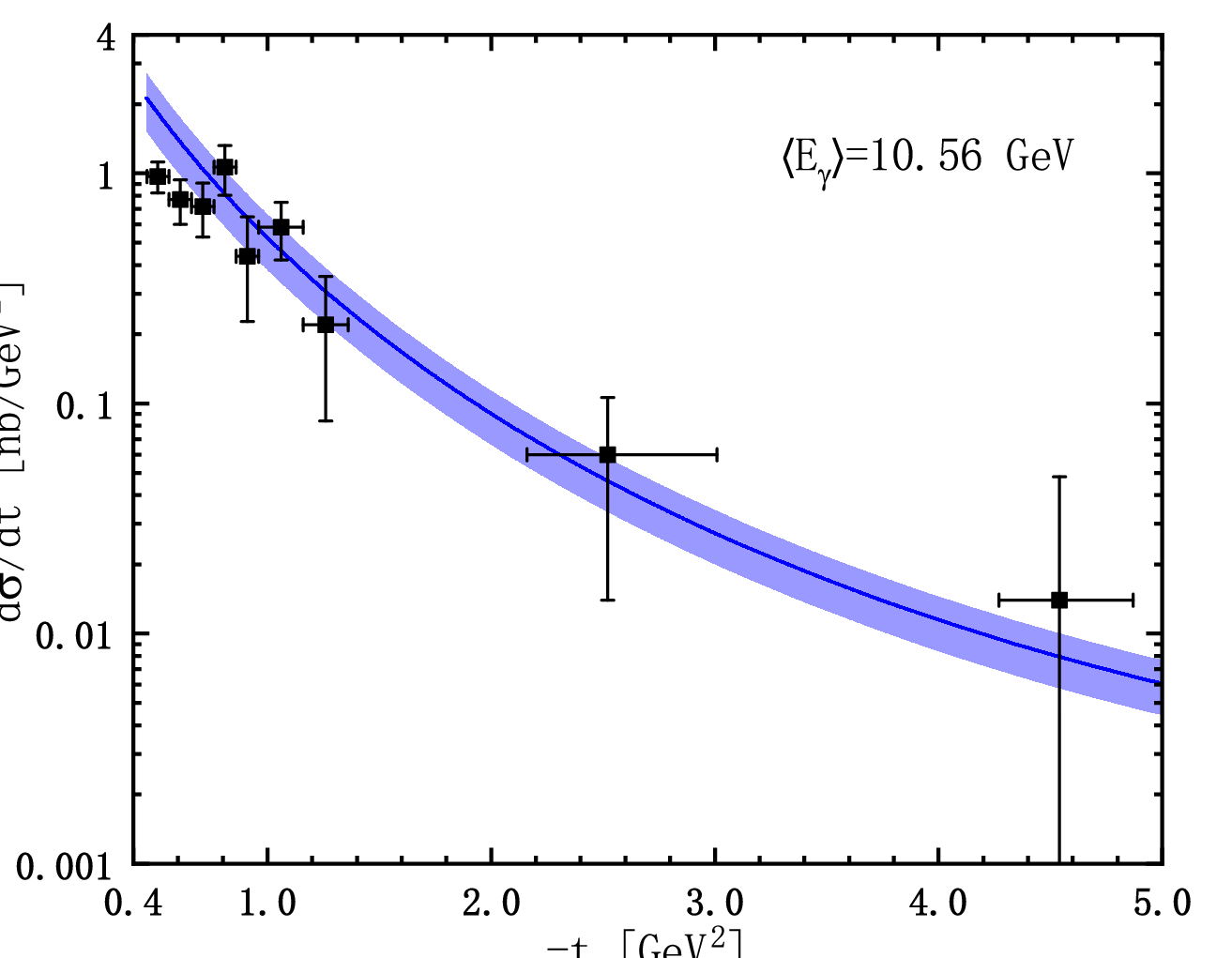}
	\includegraphics[width=0.3\columnwidth]{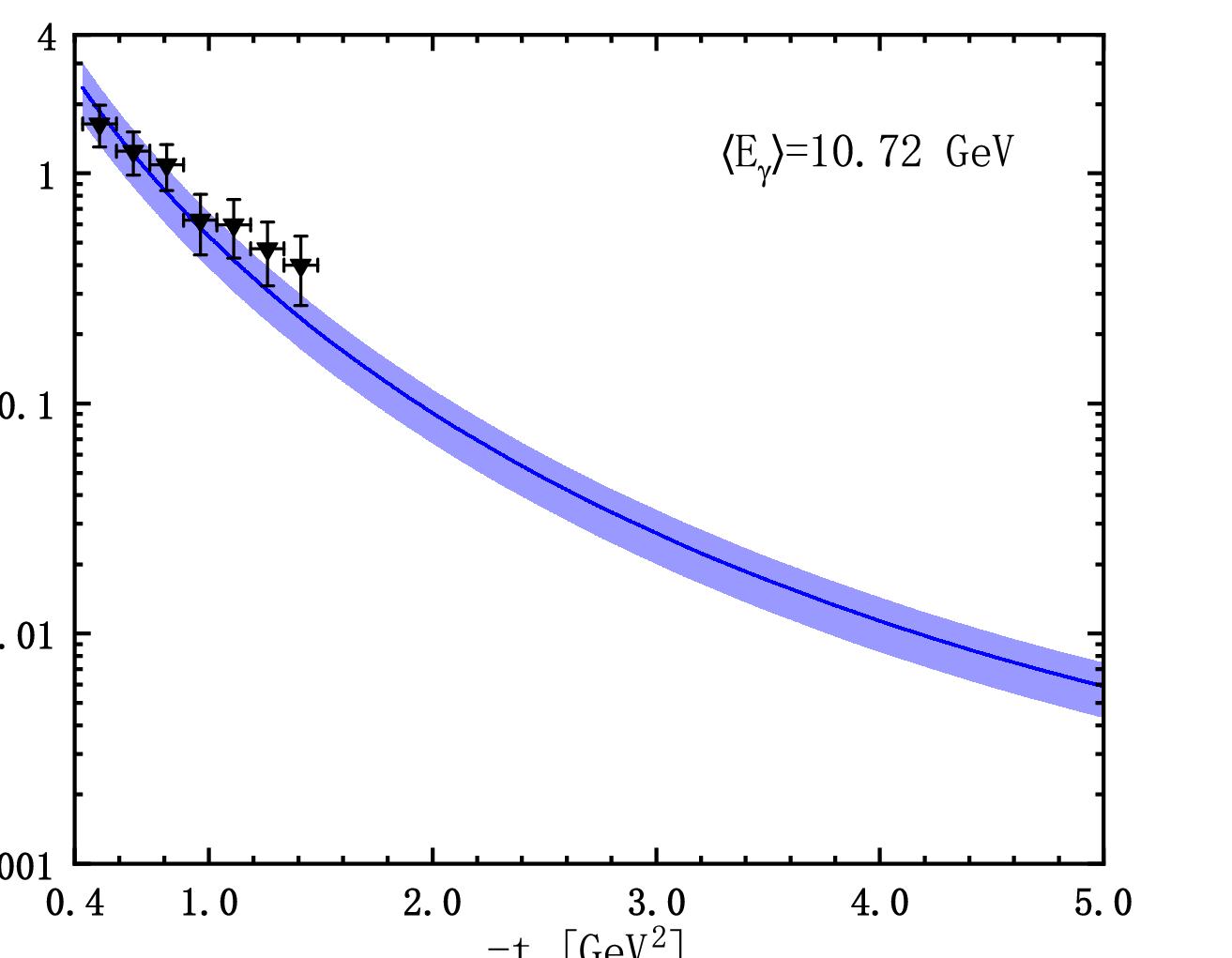}\\
	\caption{
Differential cross sections for $\gamma p \rightarrow J/\psi\,p$ as functions of $-t$ at different average photon energies $8.2\;\text{GeV}\le E_\gamma\le 11.8\;\text{GeV}$ compared with data from GlueX19~\cite{GlueX:2019mkq} (triangles), GlueX23~\cite{GlueX:2023pev} (circles), and $J/\psi$-007~\cite{Duran:2022xag} (squares) . The bands depict the parameter uncertainties.
}
	\label{fig:dif1}
\end{figure}

\begin{figure}
	\centering
	\includegraphics[width=0.45\columnwidth]{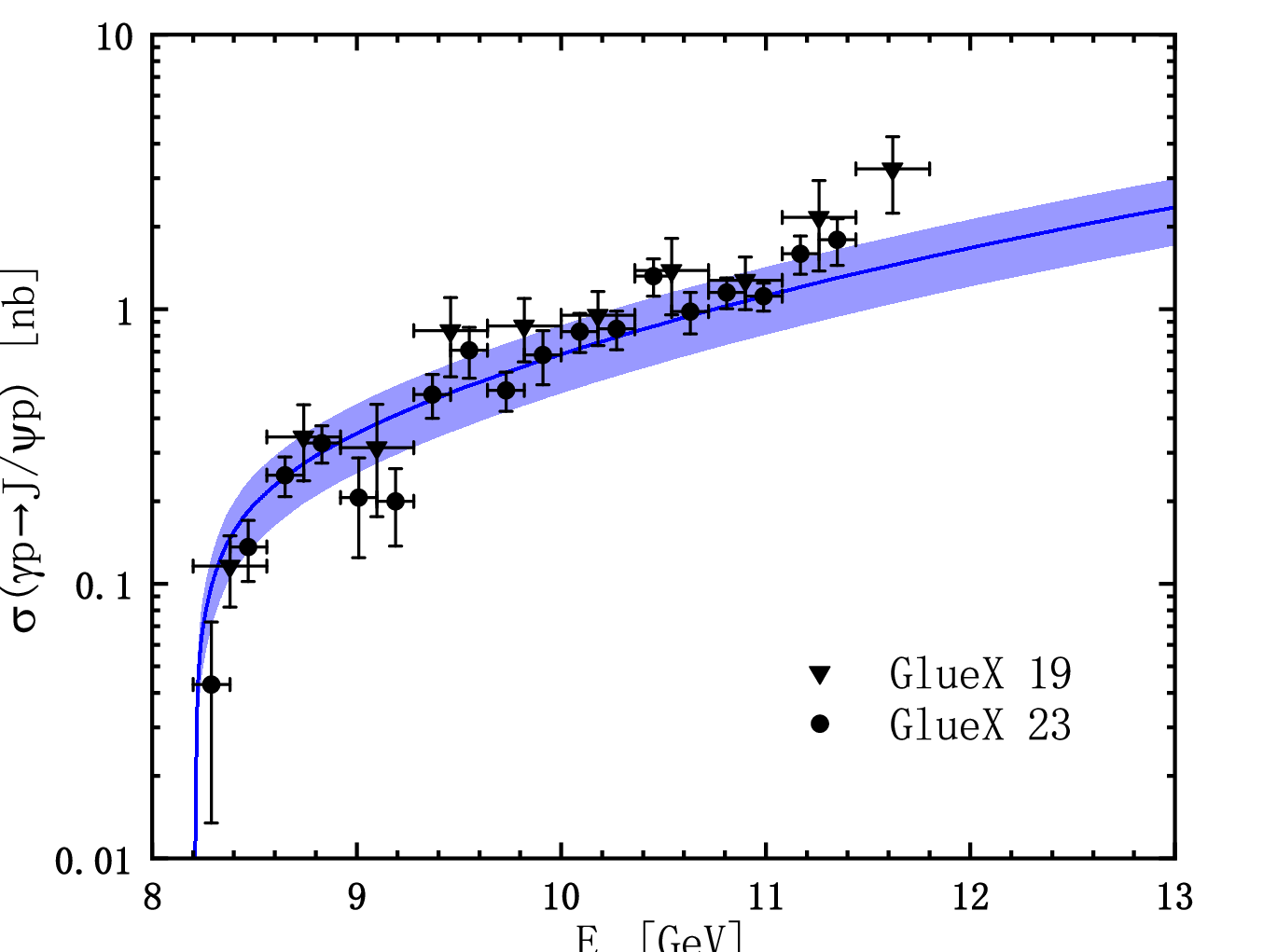}
	\caption{
Total cross section for $\gamma p \rightarrow J/\psi\,p$ as a function of $E_\gamma$ compared with data from GlueX19~\cite{GlueX:2019mkq} (triangles) and GlueX23~\cite{GlueX:2023pev} (circles). The band depicts the parameter uncertainty.
}
	\label{fig:tot1}
\end{figure}

We now present model predictions for the differential cross section (Eq.~\ref{eq:dsigma}) of near-threshold photoproduction of heavy quarkonium as a function of $t$ at various average photon energies $\langle E_\gamma\rangle$, and for the total cross section (Eq.~\ref{eq:sigma}) as a function of the laboratory photon energy $E_\gamma$.

We first analyze all available data for near-threshold $J/\psi$ photoproduction from the GlueX~\cite{GlueX:2019mkq,GlueX:2023pev} and $J/\psi$-007 experiments~\cite{Duran:2022xag}. These data include differential cross sections in 15 photon-energy bins and total cross sections, with the GlueX23 measurements on differential cross section covering the full near-threshold kinematic region. 
They have been analyzed in recent theoretical studies on gluonic properties of the proton~\cite{Hatta:2019lxo,Mamo:2019mka,Mamo:2022eui,Sun:2021pyw,Sun:2021gmi,Guo:2021ibg,Guo:2023pqw,
Guo:2023qgu,Pentchev:2025qyn,Du:2020bqj,JointPhysicsAnalysisCenter:2023qgg,Sakinah:2024cza,Tang:2024pky,
Kim:2025oyo,Lee:2022ymp,Kim:2024lxc,Mamo:2021krl,Kharzeev:2021qkd}. 
For the parameters related to the $J/\psi$ production~\cite{Guo:2021ibg}, we adopt $\alpha_\text{EM}=1/137$, $\alpha_S=0.3$, $e_c=2/3$, $M=0.938\;\text{GeV}$, $M_{J/\psi}=3.097\;\text{GeV}$, and the $J/\psi$ wave function at the origin~\cite{Eichten:1995ch,Eichten:2019hbb}
\begin{align}
	|\psi_{\text{NR}}(0)|^2={1.0952\over 4\pi}(\text{GeV})^3.
\end{align}

In Fig.~\ref{fig:dif1} we plot the model results (solid curves) for the differential cross sections as functions of $-t$ at different $\langle E_\gamma\rangle$ compared with the experimental data measured at different $-t$ bins. 
The triangle, circle, and square data points corresponds to the data from GlueX19~\cite{GlueX:2019mkq} , GlueX23~\cite{GlueX:2023pev} , and $J/\psi$-007~\cite{Duran:2022xag} , respectively.
The blue bands depict the uncertainties from the model parameters. 
Overall, the model describes the data reasonably well, with slight deviations at $\langle E_\gamma\rangle=9.86$ and $10.82\;\text{GeV}$. 
The differential cross sections exhibit the characteristic diffractive peak at forward $t$ and an exponential fall-off at larger momentum transfer. 
The GlueX23 data (circles) span the entire physical $t$ range; at the lowest average energy $\langle E_\gamma\rangle=8.93\;\text{GeV}$ an enhancement appears at the largest $|t|$, possibly signaling contributions beyond the pure $t$-channel mechanism (e.g., $u$- or $s$-channel exchanges)~\cite{JointPhysicsAnalysisCenter:2023qgg,Chudakov:2024osk}. While our model does not reproduce this enhancement at $8.93\;\text{GeV}$, similar features emerge at $\langle E_\gamma\rangle=9.86$ and $10.82\;\text{GeV}$.

In Fig.~\ref{fig:tot1}, we present the total cross section as a function of $E_\gamma$. The model prediction (curve) agrees very well with the GlueX19 and GlueX23 data.

\begin{figure}
	\centering
	\includegraphics[width=0.43\columnwidth]{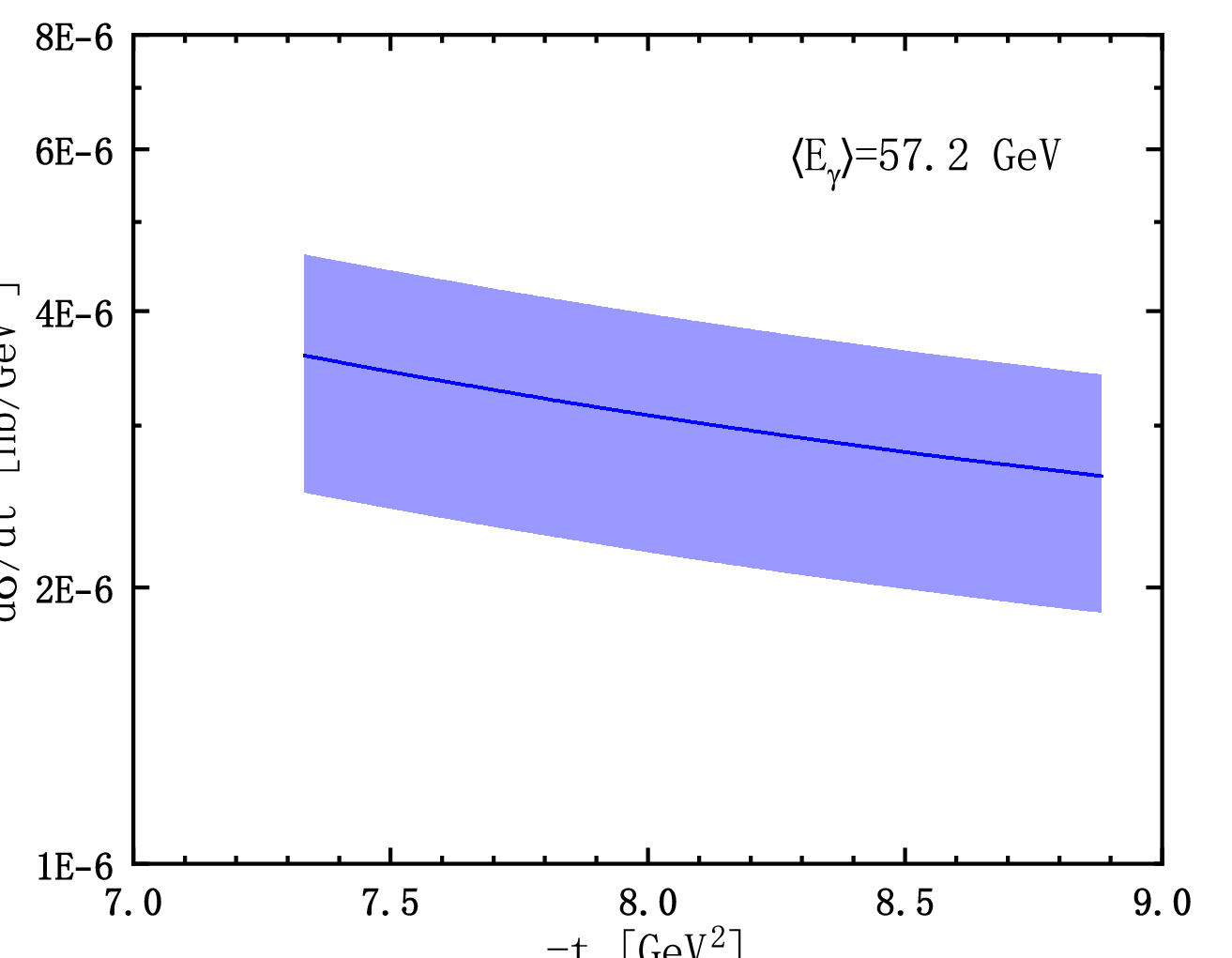}
	\includegraphics[width=0.43\columnwidth]{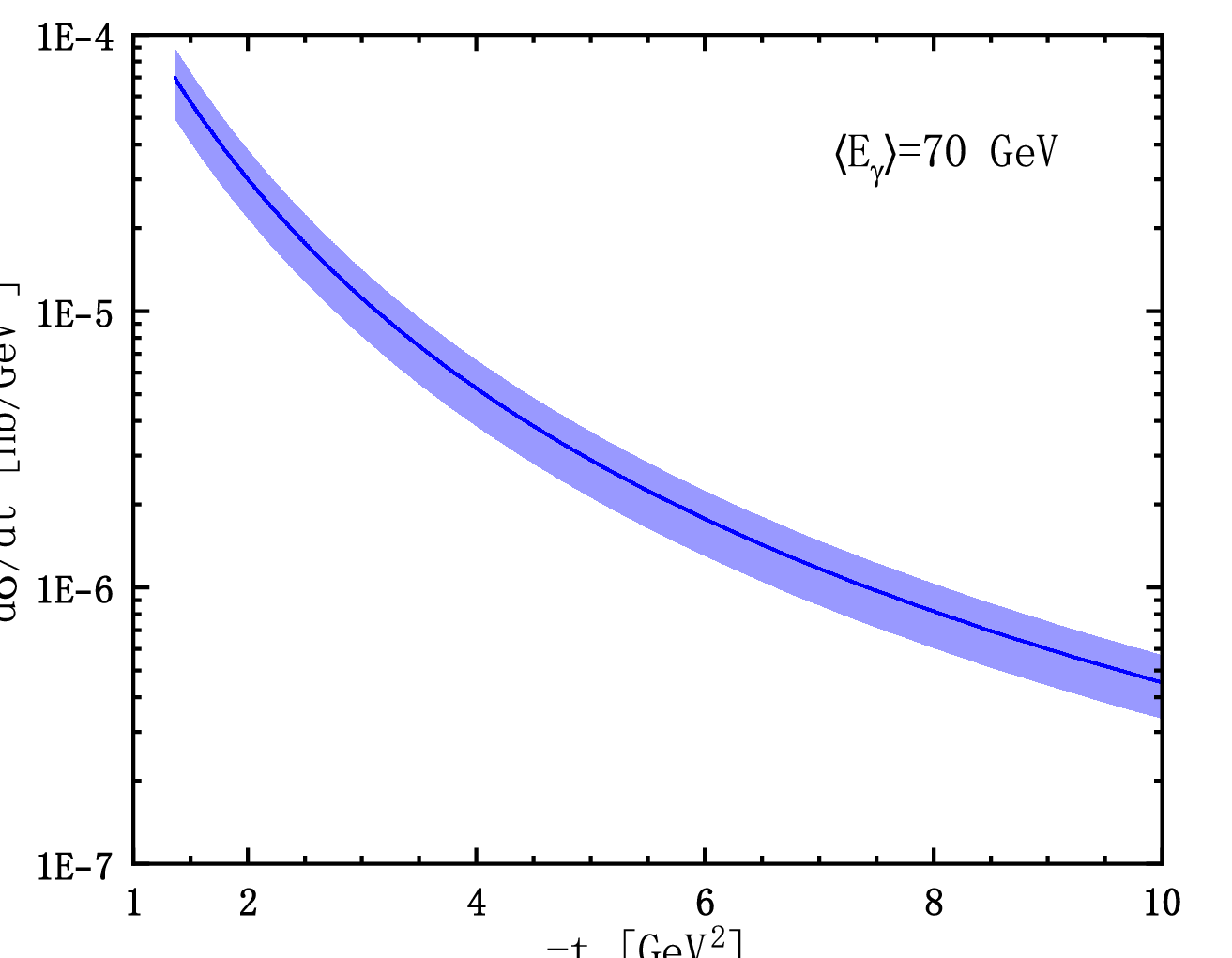}
	\caption{The predicted differential cross sections of $\gamma p\rightarrow \Upsilon p$ are plotted as functions of $-t$ at $\langle E_\gamma\rangle=57.2\,\text{GeV}$ (left panel) and $\langle E_\gamma\rangle=70\,\text{GeV}$ (right panel).}
	\label{fig:dif2}
\end{figure}

\begin{figure}
	\centering
	\includegraphics[width=0.45\columnwidth]{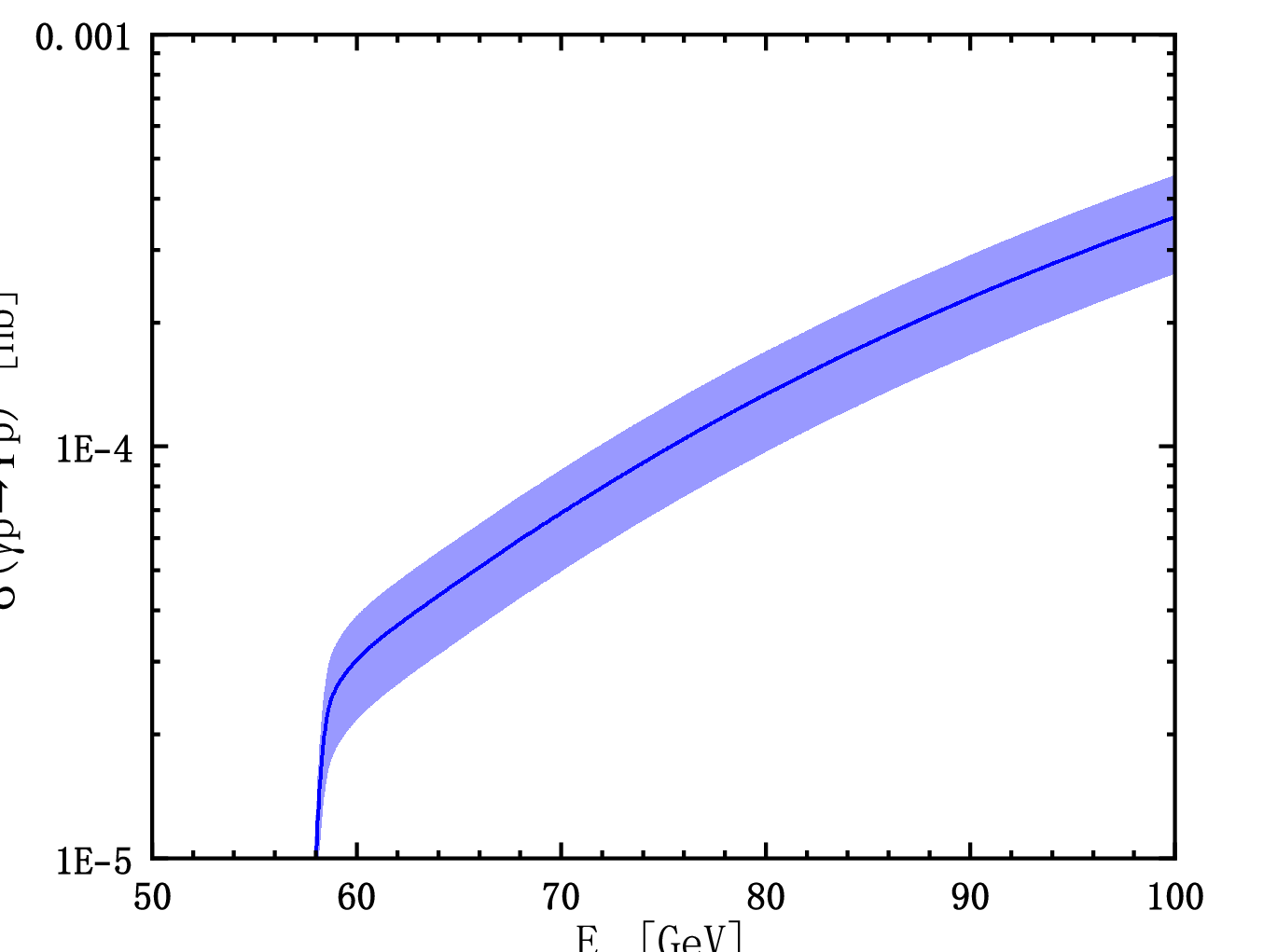}
	\caption{The predicted total cross section of $\gamma p\rightarrow \Upsilon p$ is plotted as a function of $E_\gamma$.}
	\label{fig:tot2}
\end{figure}

Finally, we present predictions for $\Upsilon$ photoproduction, using $\alpha_S=0.2$, $e_b=-1/3$, $M_\Upsilon=9.46\;\text{GeV}$, and the $\Upsilon$ wave function at the origin~\cite{Eichten:1995ch,Eichten:2019hbb}
\begin{align}
    |\psi_{\text{NR}}(0)|^2 = \frac{5.8588}{4\pi}\;\text{GeV}^3.
\end{align}
In Fig.~\ref{fig:dif2} we show the predicted differential cross sections near threshold ($\langle E_\gamma\rangle=57.2\;\text{GeV}$, left) and farther above threshold ($\langle E_\gamma\rangle=70\;\text{GeV}$, right). The total cross section is displayed in Fig.~\ref{fig:tot2}. 
The larger bottom-quark mass makes the heavy-quark limit more reliable for $\Upsilon$ production, reducing theoretical uncertainties from higher-order corrections. 
However, the production rate is strongly suppressed by the heavy quark mass, leading to cross sections much smaller than that for $J/\psi$. 
Verification of these results awaits more precise future measurements.

\section{Summary}\label{Sec:5}

In this work we studied the near-threshold exclusive photoproduction of heavy quarkonia within the GPD formalism, employing a spectator model for the proton. The amplitudes of these processes are dominated by the leading moments of gluon GPDs, offering a probe of gluon distributions at large skewness $\xi$. The model parameters were determined by simultaneously fitting $f_1^g(x)$ from the NNPDF3.1 parameterization and $g_{1L}^g(x)$ from the NNPDFpol1.1 parameterization at $\mu_0=2\;\text{GeV}$. 
The gluon GPDs $H^g(x,\xi,t)$ and $E^g(x,\xi,t)$ at nonzero $\xi$ were calculated analytically at tree level via two-gluon exchange, separately in the regions $0\le x\le\xi$ and $\xi<x\le1$.

Several gluon-related quantities were extracted and compared with recent lattice results. The gluonic contribution to the quantum anomalous energy is estimated reliably as there is a good agreement between our model and lattice data for the average gluon momentum fraction. 
The differential and total cross sections for $J/\psi$ photoproduction were computed and compared with the latest JLab data from GlueX and $J/\psi$-007 experiments, showing good overall agreement. 
Finally, we presented predictions for $\Upsilon$ photoproduction, as the heavy-quark limit is more suitable to employ here. These predictions can be tested at future EIC and EicC.

\section*{Acknowledgements}
This work is partially supported by the National Natural Science Foundation of China under grant numbers 12447136 and 12150013.

\appendix
\section{The numerators and denominators of gluon GPDs in Eqs.~(\ref{eq:GPDH}-\ref{eq:GPDE})}\label{Appendix}

The numerators in Eqs.~(\ref{eq:GPDH}-\ref{eq:GPDE}) are listed as follows:
\begin{align}
	N_H^{0 \leq  x\leq \xi}=& 4 \bm{k}_\perp\cdot \bm{\Delta}_\perp \{g_2^- g_2^+ (-1 + x)^2 \bm{\Delta}_\perp^4 (x + \xi) (x + \xi + 2 x \xi) + 4 g_2^- g_2^+ \bm{k}_\perp^2 \bm{\Delta}_\perp^2 (1 + \xi)^2 (x + \xi) [-\xi + x (-1 + 2 \xi)] \nonumber\\
	&+ 16 \bm{k}_\perp^2 M \xi^2 (1 + \xi)^2 [g_1^+ g_2^- M (x + \xi) (-1 + \xi^2) + g_2^+ (g_2^- \xi (x + \xi) (M_X + M (-1 + x) - M_X \xi)\nonumber\\
	& - g_1^- M (-1 + \xi) (x - x \xi + \xi (3 + \xi)))] - 4 M \bm{\Delta}_\perp^2 [g_1^- (1 + \xi) (4 g_1^+ M (x + \xi) (x + (-1 + x)^2 \xi - x \xi^2) \nonumber\\
	&- g_2^+ (2 M_X (x + \xi) (2 x + (-1 + x) x \xi + (1 - 3 x) \xi^2) + M (1 + \xi) (\xi^3 + x \xi (-4 + \xi + 2 \xi^2) + x^3 (6 - \xi (8 - 3 \xi)) \nonumber\\
	&- x^2 (4 + 3 \xi (-4 + \xi (2 + \xi)))))) - g_2^- (x + \xi) (g_1^+ (1 + \xi) (2 M_X (-1 + x) \xi (x + \xi) + M (-1 + \xi) (-2 x^2 - 2 x \xi \nonumber\\
	&+ (-1 + x) (1 + 3 x) \xi^2)) - g_2^+ (M (x^3 + x^2 \xi + x (-1 - 2 (-1 + x) x) \xi^2 + (-2 + x + 3 x^2 - 3 x^3) \xi^3) \nonumber\\
	&+ M_X (1 + \xi) (-\xi^3 - 2 x \xi (-1 + \xi^2) + x^2 (2 + \xi (-2 + \xi (-2 + 3 \xi))))))]+ 16 M^3 \xi^2 (x + \xi)  \nonumber\\
	&\times[g_2^+ (x - \xi) (-g_2^- \xi (x + \xi) (M + M_X - M x + M_X \xi) + g_1^- (1 + \xi) (2 M_X (-1 + x) \xi+ M (x - \xi) (1 + \xi))) \nonumber\\
	& + g_1^+ (1 + \xi) (g_2^- (x + \xi) (M (x - \xi) (-1 + \xi) + 2 M_X (-1 + x) \xi) - 4 g_1^- \xi (M (-1 + x)^2+ M_X (-1 + \xi^2)))] \}\nonumber\\
	&+\bm{\Delta}_\perp^2 \bigg\{16 g_2^- g_2^+ \bm{k}_\perp^4 (-1 + \xi) (1 + \xi)^3 (x + \xi) + 
	g_2^- g_2^+ (-1 + x)^3 (1 + x) \bm{\Delta}_\perp^4 (x + \xi) \nonumber\\
	&+ 8 g_2^- g_2^+ \bm{k}_\perp^2 (-1 + x) \bm{\Delta}_\perp^2 (1 + \xi) (x + \xi) (-3 + x \xi) - 32 \bm{k}_\perp^2 M (1 + \xi) [-g_2^- (x + \xi) (g_1^+ (-1 + \xi^2) (M_X (x + \xi) \nonumber\\
	&+ M (-x - \xi + (-3 + x) \xi^2)) + g_2^+ (-M_X (x (-1 + \xi) - \xi) \xi (-1 + \xi^2) + M (\xi^2 + 2 \xi^4 - x \xi^2 (3 + \xi) \nonumber\\
	&- x^2 (1 - \xi^2 - \xi^3)))) - g_1^- (1 - \xi^2) (2 g_1^+ M (-2 + x - \xi) (x + \xi) + g_2^+ (-M_X (x - 3 \xi) (x + \xi)+ M ((-3 + \xi) \xi^2 \nonumber\\
	& + x^2 (1 + (4 - \xi) \xi) + x \xi (-2 + \xi (7 + \xi)))))] + 8 M (-1 + x) \bm{\Delta}_\perp^2[g_2^- (x + \xi) (g_1^+ (1 + \xi) (M_X (-1 + x) (x + \xi) \nonumber\\
	& + M (1 + x) (1 - \xi) (x + (1 - x) \xi)) + g_2^+ ( M (1 - x + x^3) \xi^2-M x^2  - M_X (1 + x) \xi (1 + \xi) (-x - (1 - x) \xi)))\nonumber\\
	& - g_1^- (1 + \xi) (2 g_1^+ M (x + \xi) (2 + x^2 - \xi - x (1 + \xi))+ g_2^+ (-M_X (x (-1 + x - 3 \xi) - \xi) (x + \xi) - M (1 + x)\nonumber\\
	& \times (x^2 (-1 + \xi)^2 + \xi^2 (1 + \xi) - x \xi (-2 + \xi (3 + \xi)))))]+ 16 M^3 (x + \xi) [g_2^- (x + \xi) (2 g_1^+ (1 + \xi) (-M_X x^2 \nonumber\\
	& + M_X (1 - 2 (1 - x) x) \xi^2 + 
	M (x - \xi) (1 - \xi) (x + (1 - x) \xi))+ g_2^+ (x - \xi) (-M x^2 + M (1 + 2 (-1 + x) x) \xi^2\nonumber\\
	&  + 2 M_X \xi (x + \xi + \xi^2 - x \xi^2)))- 2 g_1^- (1 + \xi) (-g_2^+ (x - \xi) ((M - M_X) x^2 - 2 (M_X + M (-1 + x)) x \xi \nonumber\\
	&  + (M (1 - (3 - x) x) + M_X (-1 - 2 (1 - x) x)) \xi^2 + M (1 - x) \xi^3) + 2 g_1^+ (2 M_X (-1 + \xi^2) (x^2 + (-1 + x) \xi^2)\nonumber\\
	& + M ((2 - x) x^2- x^2 \xi + ( x (3 - 2 (2 - x) x)-2) \xi^2 + \xi^3)))] + 16 (1 + \xi) [4 g_1^- g_2^+ M^2 (1 - x) (x - \xi) \xi^2 (1 - \xi^2) \nonumber\\		
	&+ g_2^- (x + \xi) (4 g_1^+ M^2 (1 + x) \xi^2 (-1 + \xi^2) + g_2^+ ((-1 + x) \bm{\Delta}_\perp^2 (1 + x \xi) + 4 M \xi^2 ((M + M_X) x - M (1 - x) x \xi \nonumber\\
	&- (M + M_X x) \xi^2)))] \frac{(\bm{k}_\perp\cdot\bm{\Delta}_\perp)^2}{\bm{\Delta}_\perp^2}\bigg\},
	\tag{A1}\\
	N_H^{\xi< x \leq 1}=&8 \bm{k}_\perp\cdot \bm{\Delta}_\perp \{g_2^- g_2^+ (-1 + x)^3 \bm{\Delta}_\perp^4 \xi + 4 g_2^- g_2^+ \bm{k}_\perp^2 (1 - x) \bm{\Delta}_\perp^2 \xi (1 - \xi^2)+4 \bm{k}_\perp^2 M \xi^2 (-1 + \xi^2)   \nonumber\\
	& \times[g_1^+ g_2^- M (1 + \xi) (-2 + x + \xi)+ g_2^+ (g_1^- M (-2 + x - \xi) (-1 + \xi)- g_2^- (M - M_X) (1 - x) \xi)] - (1 - x) \bm{\Delta}_\perp^2\nonumber\\
	&  \times[g_2^+ (g_2^- \xi (-M_X^2 (2 + x^2 - 3 \xi^2) + M M_X (2 (-2 + x) x+ (1 + (4 - 3 x) x) \xi^2)+ M^2 ((2 - x (4 - 3 x)) \xi^2  +2\nonumber\\
	&  - 4 x + x^2 )) + g_1^- M (2 M_X (x + \xi - (1 - x) x \xi+ (1 - 2 x) \xi^2 - \xi^3) + M (1 + \xi) (2 (-1 + x) x +x (-5 + 3 x) \xi^2 \nonumber\\
	& - 2 (1 - 2 (2 - x) x) \xi + 3 (1 - x) \xi^3)))+ g_1^+ M (4 g_1^- M \xi (-2 + (2 - x) x + \xi^2)+ g_2^- (2 M_X (\xi + x^2 \xi -\xi^2 (1 + \xi)\nonumber\\
	&  + x (-1 + \xi) (1 + 2 \xi)) + M (-1 + \xi) (2 \xi - 3 \xi^3 + x (1 + 3 \xi) (-2 + (-2 + \xi) \xi) + x^2 (2 + \xi (4 + 3 \xi)))))]\nonumber\\
	&+ 4 M^2 \xi^2 [g_2^+ (x - \xi) (-g_2^- M_X (M + M_X) (-1 + x) \xi (x + \xi) + g_1^- (2 M M_X (-1 + x)^2 \xi + M^2 (-1 + x)^2 (1 + \xi) \nonumber\\
	&- M_X^2 (1 - \xi) (1 + \xi)^2)) + g_1^+ (g_2^- (x + \xi) (M^2 (1 - x)^2 (-1 + \xi) + 2 M M_X (1 - x)^2 \xi+ M_X^2 (1 - \xi)^2 (1 + \xi)) \nonumber\\
	& + 4 g_1^- M (1 - x) \xi (M (1 - x)^2 - M_X (1 - \xi^2)))]\}+\bm{\Delta}_\perp^2 \bigg\{16 g_2^- g_2^+ \bm{k}_\perp^4 (1 - \xi^2)^2 + g_2^- g_2^+ (1 - x)^4 \bm{\Delta}_\perp^4 \nonumber\\
	& - 8 g_2^- g_2^+ \bm{k}_\perp^2 (1 - x)^2 \bm{\Delta}_\perp^2 (1 - \xi^2) + 8\bm{k}_\perp^2 (1 - \xi^2) [2 g_1^- M (2 g_1^+ M (2- \xi^2+ (-2 + x) x ) + g_2^+ (-(M + M_X) x^2 \nonumber\\
	& - 2 (M_X - M (1 - x)) (1 - x) \xi + (M_X - M (3 + (-5 + x) x)) \xi^2 + M (-1 + x) \xi^3)) + g_2^- (g_2^+ (x (2 M M_X x\nonumber\\
	& + M_X^2 (2 + x) + M^2 (-2 + 3 x)) + (-3 M_X^2 + 2 M M_X (1 - (3 - x) x) + M^2 (3 - 2 x (1 + x))) \xi^2)\nonumber\\
	& - 2 g_1^+ M ((M + M_X) x^2 + 2 (M_X - M (1 - x)) (-1 + x) \xi + ( M (3 - (5 - x) x)-M_X ) \xi^2 + M (-1 + x) \xi^3))]\nonumber\\
	& - 2 (1 - x)^2 \bm{\Delta}_\perp^2 [2 g_1^- M (2 g_1^+ M (2 - (2 - x) x - \xi^2) + g_2^+ ((M_X-M ) (2 - x) x + (M_X - M (1 - (3 - x) x)) \xi^2 \nonumber\\	
	& + 2 M_X x \xi - M (1 - x) \xi^3)) - g_2^- (g_2^+ x ( M_X^2 (2 - x) + 2 M M_X x-M^2 (2 - x) )+ g_2^+ (M^2 (1 - 2 (1 - x) x)  \nonumber
\end{align}
\begin{align}		
	&+3 M_X^2 + 2 M M_X (1 + x - x^2)) \xi^2 + 2 g_1^+ M (M_X ((-2 + x) x + 2 x \xi - \xi^2)+ M (-1 + \xi) ((-2 + x) x  \nonumber\\		
	& - (2 - x) x \xi- (1 - x) \xi^2)))]- 8 [-g_2^- (x + \xi) (g_2^+ (M + M_X) (x - \xi) ((M + M_X) (M_X + M (-1 + x))^2 \nonumber\\
    & - M_X (M_X^2 + 2 M^2 (1 - x)^2- M M_X (1 - 2 x)) \xi^2) + 2 g_1^+ M (M^3 (1 - x)^2 (-1 + \xi) (x - (1 - x) \xi) \nonumber\\
    & - M_X^3 (x - \xi) (1 - \xi^2) + M M_X^2 (-1 + \xi^2) (x (-1 + 2 x) + \xi - 2 x \xi - (1 - x) \xi^2)- M^2 M_X (1 - x) \nonumber\\
    & \times(\xi + x (-1- x + \xi - 2 (1 - x) \xi^2)))) + 2 g_1^- M (2 g_1^+ M (M_X^2 (\xi^2-x^2 ) (1 - \xi^2) + 2 M M_X (1 - x) (1 - \xi^2)\nonumber\\
    & \times (x^2 - (1 - x) \xi^2)+ M^2 (1 - x)^2 ( (1 - 2 (1 - x) x) \xi^2-x^2 )) - g_2^+ (x - \xi) (M^3 (1 - x)^2 (x (\xi-1 ) - \xi) (1 + \xi)\nonumber\\
    &+ M_X^3 (x + \xi) (-1 + \xi^2) + M^2 M_X (x - x^3 + \xi - x^2 \xi + 2 (-1 + x)^2 x \xi^2) + M M_X^2 (-1 + \xi^2) (2 x^2 - \xi (1 + \xi) \nonumber\\
    &+ x (-1 + \xi (2 + \xi)))))] + 16 (-1 + x) \xi^2 [2 g_1^+ g_2^- M^2 (x + \xi) (-1 + \xi^2) + g_2^+ (2 g_1^- M^2 (x - \xi) (-1 + \xi^2) \nonumber\\
    &+ g_2^- ((-1 + x) \bm{\Delta}_\perp^2 - 2 M M_X x (-1 + \xi^2) + 2 M^2 (x + (-2 + x) \xi^2)))] \frac{(\bm{k}_\perp\cdot\bm{\Delta}_\perp)^2}{\bm{\Delta}_\perp^2}\bigg\} ,
	\tag{A2}\\
	N_E^{0 \leq  x\leq \xi}=&2 \bm{k}_\perp\cdot \bm{\Delta}_\perp \{4 \bm{k}_\perp^2 (1 + \xi)^2 [g_1^+ g_2^- M (x + \xi) (\xi^2-1 ) + g_2^+ (g_2^- \xi (x + \xi) (M_X - M (1 - x) - M_X \xi) + g_1^- M (1 - \xi)\nonumber\\
	&\times (x - x \xi + \xi (3 + \xi)))] - (1 - x) \bm{\Delta}_\perp^2 [g_1^+ g_2^- M (1 + 3 x) (x + \xi) (-1 + \xi^2) + g_2^+ (-g_1^- M (1 + \xi) (x^2 (5 - 3 \xi) \nonumber\\
	&+ \xi + \xi^2 + x (3 \xi (2 + \xi)-5 )) - 
	g_2^- (x + \xi) (M_X (1 + \xi) ( \xi + 3 x \xi-2 x ) +M (\xi - x^2 (2 + 3 \xi))))]+ 4 M^2 (x + \xi) \nonumber\\
	&\times [g_2^+ (x - \xi) (-g_2^- \xi (x + \xi) (M + M_X - M x + M_X \xi) + g_1^- (1 + \xi) (2 M_X (-1 + x) \xi + M (x - \xi) (1 + \xi)))\nonumber\\
	& + g_1^+ (1 + \xi) (g_2^- (x + \xi) (M (x - \xi) (-1 + \xi) + 2 M_X (-1 + x) \xi) - 4 g_1^- \xi (M (-1 + x)^2 + M_X (-1 + \xi^2)))]\}\nonumber\\
	& + \bm{\Delta}_\perp^2 \bigg\{4 \bm{k}_\perp^2 (1 + \xi) [g_1^+ g_2^- M (3 - x) (x + \xi) (1 - \xi^2) + g_2^+ (g_2^- (x + \xi) (M (-3 + x) x- (1 - x) (M_X + M x) \xi \nonumber\\
	& + (2 M + M_X - M_X x) \xi^2) + g_1^- M ((3 - x) x + \xi + (5 - 
	4 x) x \xi + (4 - (7 - x) x) \xi^2- (1 + x) \xi^3))] \nonumber\\
	&  + (1 - x)^2 (1 + x) \bm{\Delta}_\perp^2 [g_1^+ g_2^- M (-1 + \xi) (x + \xi) + g_2^+ (g_2^- (x + \xi) (M x - M_X \xi) + g_1^- M (x (-1 + \xi) \nonumber\\
	& - \xi (3 + \xi)))]+ 4 M^2 (-1 + x) (x + \xi) [g_2^- (x + \xi) (g_2^+ (x - \xi) (M x - M_X \xi)+ g_1^+ (M (x - \xi) (-1 + \xi) \nonumber\\
	& + 2 M_X x (1 + \xi))) + g_1^- (-4 g_1^+ (1 + \xi) ( M_X (x + \xi^2)-M (1 - x) x )+ g_2^+ (x - \xi) (2 M_X x (1 + \xi)- M (x (1 - \xi)   \nonumber\\	
	&+ \xi (3 + \xi))))] + 8 (1 + \xi) [g_1^+ g_2^- M (1 + x) (x + \xi) (-1 + \xi^2)+g_2^+ (g_1^- M (-1 + x) (x - \xi) (-1 + \xi^2)\nonumber\\
	&  + g_2^- (x + \xi) ((M + M_X) x + M (-1 + x) x \xi - (M + M_X x) \xi^2))] \frac{(\bm{k}_\perp\cdot\bm{\Delta}_\perp)^2}{\bm{\Delta}_\perp^2}\bigg\} ,
	\tag{A3}\\			
	N_E^{\xi< x \leq 1}=&2 \bm{k}_\perp\cdot \bm{\Delta}_\perp \{4 \bm{k}_\perp^2 ( \xi^2-1 ) [g_1^+ g_2^- M (1 + \xi) ( x + \xi-2 ) + g_2^+ (g_1^- M (2 - x + \xi) (1 - \xi) - g_2^- (M - M_X) (1 - x) \xi)]\nonumber\\
	& + (1 - x)^2 \bm{\Delta}_\perp^2 [g_1^+ g_2^- M (-2 + x + \xi) (1 + 3 \xi) + g_2^+ (g_2^- (M - M_X) (-1 + 3 x) \xi + g_1^- M (2 - x + \xi) (1 - 3 \xi))] \nonumber\\
	&+ 4 M [g_2^+ (x - \xi) (-g_2^- M_X (M + M_X) (-1 + x) \xi (x + \xi) + g_1^- (2 M M_X (-1 + x)^2 \xi + M^2 (-1 + x)^2 (1 + \xi) \nonumber\\
	&+ M_X^2 (-1 + \xi) (1 + \xi)^2)) + g_1^+ (g_2^- (x + \xi) (M^2 (1 - x)^2 (-1 + \xi) + 2 M M_X (1 - x)^2 \xi + M_X^2 (1 - \xi)^2 (1 + \xi)) \nonumber\\
	&+ 4 g_1^- M (1 - x) \xi (M (1 - x)^2 - M_X (1 - \xi^2)))]\} - (1 - x) \bm{\Delta}_\perp^2 \bigg\{4 \bm{k}_\perp^2 [g_1^+ g_2^- M (1 + \xi) (2 + 
	x - (5 - x) \xi + \xi^2) \nonumber\\
	&+ g_2^+ (-g_2^- (3 M + M_X) x + g_2^- (-M_X (-2 + x) + M (2 + x)) \xi^2 - g_1^- M (-1 + \xi) (2 + x - x \xi + \xi (5 + \xi)))] \nonumber\\
	&+ (1 - x)^2 \bm{\Delta}_\perp^2 [g_2^+ (g_2^- (M - M_X) x - g_1^- M (2 - x + \xi)) - g_1^+ g_2^- M (2 - x - \xi)] \nonumber\\
	&- 4 [-g_2^- (x + \xi) (-g_2^+ M_X (M + M_X) (M_X - M (1 - x)) (x - \xi) + g_1^+ M (M^2 (-1 + x)^2 + 2 M M_X (-1 + x) x \nonumber\\
	&+ M_X^2 ( 2 x-1  - (2 - \xi) \xi))) + g_1^- M (4 g_1^+ M (-1 + x) ( M_X (x + \xi^2)-M (1 - x) x ) - g_2^+ (x - \xi) (M^2 (1 - x)^2 \nonumber\\
	&+ 2 M M_X (-1 + x) x + M_X^2 (-1 + 2 x + \xi (2 + \xi))))] + 8 [g_1^+ g_2^- M (x + \xi) (-1 + \xi^2) + g_2^+ (g_2^- (M + M_X) x \nonumber\\
	&+ g_2^- (M (-2 + x) - M_X x) \xi^2 + g_1^- M (x - \xi) (-1 + \xi^2))] \frac{(\bm{k}_\perp\cdot\bm{\Delta}_\perp)^2}{\bm{\Delta}_\perp^2}\bigg\} .
	\tag{A4}
\end{align}	
The denominators are given by
\begin{align}
	D_\text{SM}^{0 \leq  x\leq \xi}=& \{4 (x + \xi) [M^2 (-1 + x) + 
	M_X^2 (1 + \xi)] + [(-1 + x) \bm{\Delta}_\perp + 
	2 \bm{k}_\perp (1 + \xi)]^2\} \nonumber\\
	&\times\{\xi [(-1 + x^2) \bm{\Delta}_\perp^2 + 
	4 M^2 (x^2 - \xi^2) + 4 \bm{k}_\perp^2 (-1 + \xi^2)] + 
	4 x(-1 + \xi^2) \bm{k}_\perp \cdot\bm{\Delta}_\perp \},
	\tag{A5}\\
	D_\text{SM}^{\xi< x \leq 1}=& \{4 (x - \xi) [M^2 (-1 + x) - 
	M_X^2 (-1 + \xi)] + [(-1 + x) \bm{\Delta}_\perp + 
	2 \bm{k}_\perp (-1 + \xi)]^2\}\nonumber\\
	&\times \{4 (x + \xi) [M^2 (-1 + x) + 
	M_X^2 (1 + \xi)] + [(-1 + x) \bm{\Delta}_\perp + 
	2 \bm{k}_\perp (1 + \xi)]^2\}. 
	\tag{A6}
\end{align}

\end{document}